\documentclass{aastex63}
\usepackage{amsmath,amstext}
\usepackage{bm}
\usepackage{color}
\usepackage{ulem}

\received{November 19, 2020}
\accepted{June 25, 2021}
\submitjournal{ApJ}

\shorttitle{Accretion Shocks around Proto-gas Giant}
\shortauthors{Takasao et al.}
\graphicspath{{./}{figures/}}

\begin{document}

\title{Hydrodynamic Model of H$\alpha$ Emission from Accretion Shocks \\
of a Proto-Giant Planet and Circumplanetary Disk}

\correspondingauthor{Shinsuke Takasao}
\email{takasao@astro-osaka.jp}

\author[0000-0003-3882-3945]{Shinsuke Takasao}
\affiliation{Department of Earth and Space Science, Graduate School of Science, Osaka University, Toyonaka, Osaka 560-0043, Japan}

\author[0000-0003-0568-9225]{Yuhiko Aoyama}
\affiliation{Institute for Advanced Study, Tsinghua University, Beijing 100084, People's Republic of China}
\affiliation{Department of Astronomy, Tsinghua University, Beijing 100084, People's Republic of China}


\author[0000-0002-5658-5971]{Masahiro Ikoma}
\affiliation{Division of Science, National Astronomical Observatory of Japan, Mitaka, Tokyo 181-8588, Japan}
\affiliation{Department of Astronomical Science, The Graduate University for Advanced Studies, SOKENDAI, Mitaka, Tokyo 181-8588, Japan} 
\affiliation{Department of Earth and Planetary Science, Graduate School of Science, The University of Tokyo, Bunkyo-ku, Tokyo 113-0033, Japan}
 
\begin{abstract}
Recent observations have detected excess H$\alpha$ emission from young stellar systems with an age of several Myr such as PDS 70. One-dimensional radiation-hydrodynamic models of shock-heated flows that we developed previously demonstrate that planetary accretion flows of $>$ a few ten km s$^{-1}$ can produce H$\alpha$ emission. It is, however, a challenge to understand the accretion process of proto-giant planets from observations of such shock-originated emission because of a huge gap in scale between the circumplanetary disk (CPD) and the microscopic accretion shock. To overcome the scale gap problem, we combine two-dimensional, high-spatial-resolution global hydrodynamic simulations and the one-dimensional local radiation-hydrodynamic model of the shock-heated flow. From such combined simulations for the protoplanet-CPD system, we find that the H$\alpha$ emission is mainly produced in localized areas on the protoplanetary surface. The accretion shocks above the CPD produce much weaker H$\alpha$ emission (approximately one to two orders of magnitude smaller in luminosity). Nevertheless, the accretion shocks above the CPD significantly affect the accretion process onto the protoplanet. The accretion occurs at a quasi-steady rate if averaged on a 10 day timescale, but its rate shows variability on shorter timescales. The disk surface accretion layers including the CPD shocks largely fluctuate, which results in the time-variable accretion rate and H$\alpha$ luminosity of the protoplanet. We also model the spectral emission profile of the H$\alpha$ line and find that the line profile is less time-variable despite the large variability in luminosity. High-spectral-resolution spectroscopic observation and monitoring will be key to revealing the property of the accretion process.
\end{abstract}



\section{Introduction} \label{sec:intro}
Revealing the formation mechanism of gas giant planets is highly important for advancing our understanding of 
planetary system formation because such massive planets have a significant impact on the formation processes of other planets and the final architecture of planetary systems \citep[e.g.,][]{2011Natur.475..206W}. 
The most widely accepted scenario for the formation of gas giant planets is the core accretion scenario in which a solid core with a mass larger than a critical value acquires a massive gas envelope from the protoplanetary disk via a runaway gas accretion process 
\citep{1980PThPh..64..544M,1986Icar...67..391B}.
While the early-phase gas accretion proceeds by the Kelvin-Helmholtz contraction of the proto-envelope \citep{2000ApJ...537.1013I}, the late-phase gas accretion is controlled by the gas supply from the protoplanetary disk \citep{2002ApJ...580..506T,2007ApJ...667..557T}.
Although we have such a general picture of the giant planet formation, the details of how proto-giant planets receive mass from their surroundings still remain unclear. As the accretion process occurring near protoplanets ultimately determines the subsequent evolution of the planets by regulating the injection of mass, heat, and angular momentum, it is important to reveal the accretion process including the flows and shocks in the vicinity of protoplanets.

Observational studies have been rapidly progressing since the report of the detection of H$\alpha$ emission from the protoplanet candidate LkCa~15~b \citep{2015Natur.527..342S}, although the origin of the H$\alpha$ excess remains controversial \citep[e.g.,][]{2019ApJ...877L...3C}. PDS~70b is known as the first example of a young planet that was robustly detected in H$\alpha$ \citep{2018ApJ...863L...8W,2019NatAs...3..749H,2020AJ....159..222H}. The H$\alpha$ emission is believed to be produced by accretion-heated gas near the protoplanet. The mass of PDS~70b estimated with different methods ranges approximately from a few to 17~$M_{\rm J}$ \citep[for instance, see Table~1 of][]{2019ApJ...885L..29A}. If the actual mass is within this range, the protoplanet is probably a proto-giant planet in the late phase of gas accretion. The observed H$\alpha$ luminosity is $\sim 10^{26}$-$10^{27}$~erg~s$^{-1}$, and the mass accretion rate is estimated to be $\sim 10^{-8}~M_{\rm J}~{\rm yr^{-1}}$ \citep{2018ApJ...863L...8W,2019NatAs...3..749H,2019ApJ...885L..29A,2019ApJ...885...94T,2020AJ....159..222H}. In addition, multiepoch observations with different instruments (the Multi Unit Spectroscopic Explorer (MUSE; \citet{2010SPIE.7735E..08B}) and the Magellan Adaptive Optics System (MagAO; \citet{2012SPIE.8447E..0XC})) yielded different values of the H$\alpha$ luminosity. This suggests that the accretion may be time-variable \citep{2018ApJ...863L...8W,2019NatAs...3..749H,2020AJ....159..222H}, although the influence of the observations with different instruments should be taken into account.
Furthermore, spectroscopic observations found a H$\alpha$ line width of $100$-$200~{\rm km~s^{-1}}$ for PDS 70b \citep{2019NatAs...3..749H,2020AJ....159..222H}, which is comparable to  the escape velocity at the protoplanetary surface ($\sim 150~{\rm km~s^{-1}}$) for a protoplanet with a mass of $12M_{\rm J}$ and a radius of $2R_{\rm J}$.
We note that, as pointed out by \citet{2019ApJ...885...94T}, the spectral resolution of MUSE is $\sim$120~km~s$^{-1}$ around the H$\alpha$ line center and the measured line widths should be considered to be upper limits.

By analogy with stellar accretion \citep[e.g.,][]{2016ARA&A..54..135H},
it is commonly considered that accretion shocks produce H$\alpha$-emitting gas. 
Planetary accretion is, however, geometrically and energetically different from stellar accretion and there is no consensus about the property of accretion shocks relevant to the observed H$\alpha$ emission.
Three-dimensional (3D) simulations of the accretion flow from the protoplanetary (or circumstellar) disk toward the protoplanet and the circumplanetary disk (CPD) have been performed to study the accretion structure. They found that the protoplanetary disk gas flows into high latitudes of the protoplanetary Hill sphere and the accretion toward the protoplanet mainly occurs outside the CPD or in the narrow surface layers of the CPD \citep[see Figure~\ref{fig:ic_density};][]{2008ApJ...685.1220M,2012ApJ...747...47T,2013ApJ...779...59G,2017MNRAS.465L..64S}. The vertical accretion from high latitudes can produce accretion shocks at the CPD surface \citep{2012ApJ...747...47T}, and the amount of the gravitational energy released at the shock will be sufficient to be observable \citep{,2017MNRAS.465L..64S}. As recent observations suggest the existence of a CPD around PDS~70b \citep{2018A&A...617L...2M,2019ApJ...879L..25I,2019ApJ...877L..33C}, the role of CPD for production of H$\alpha$ emission should be investigated. Also, H$\alpha$ emission may arise from the accretion shocks formed well above the protoplanet and CPD ($>$several tens of $R_{\rm J}$), depending on the size of the protoplanet and the thermodynamics around it \citep{2020arXiv200209918S}.

Gas flow accreting directly onto the central protoplanet is thought to contribute to the H$\alpha$ emission.
\citet{2019ApJ...885L..29A} applied their radiation-hydrodynamic model of shock-heated accretion flows to PDS~70b, and found that accretion shocks at the protoplanetary surface with the filling factor of $\sim 10^{-3}$ can reproduce the H$\alpha$ emission that is consistent in luminosity and line width with the observational data from \citet{2019NatAs...3..749H} \citep[see also][]{2020AJ....159..222H}; the observed upper limit for the 10\% line width of the H$\alpha$ line is $\sim$200~km~s$^{-1}$, and the line width of $\gtrsim$100~km~s$^{-1}$ can be explained theoretically by the spectral broadening in the hot ($\sim 10^{6}$~K) postshock region. 
As in the case of the accretion of T Tauri stars, magnetospheric accretion may be taking place around PDS~70b, where a convergent accretion flow is expected. However, there is no clear evidence that proto-giant planets including PDS~70b possess magnetic fields strong enough to drive magnetospheric accretion. Also, it remains unclear if the inner CPD is sufficiently ionized to couple with the protoplanetary magnetic fields \citep[we note the theoretical estimation by][]{2018AJ....155..178B}. Therefore, it is still important to study the hydrodynamic accretion process just around protoplanets.

Motivated by the observations of PDS~70b, we investigate the accretion process occurring in the vicinity of a proto-giant planet surrounded by a CPD.
Connecting a huge gap in scale between the CPD and the microscopic accretion shock is required to reveal the accretion process from observations of shock-originated emissions.
Therefore, we combine global hydrodynamic simulations with the local radiation-hydrodynamic model of shock-heated accretion flows developed by \citet{2018ApJ...866...84A}. 
In this paper we will report our first attempt to connect such a huge gap in between the micro and macro scales.
The H$\alpha$ emission predicted from our model will be described. A brief comparison with observations of PDS~70b will also be presented.

\section{Numerical Setup} \label{sec:setup}
\subsection{Method and Model Description}
We study the accretion process in the vicinity of a proto-giant planet surrounded by a CPD, using axisymmetric 2D hydrodynamic simulations. We solve the hydrodynamic equations in a conservative form using Athena++ \citep{2020ApJS..249....4S}. We use the second-order piecewise linear reconstruction method and the Harten--Lax--van
Leer Contact (HLLC) approximate Riemann solver. The equations are integrated using the third-order accurate Strong Stability Preserving Runge-Kutta method. 
We adopt the equation of state for an ideal gas, $p \propto \rho T$, where $p$, $\rho$, and $T$ are the gas pressure, density, and temperature, respectively.  
The internal energy density $e_{\rm int}$ is written as $e_{\rm int}=p/(\gamma-1)$, where $\gamma$ is the effective adiabatic index.

In this study, to take into account the effects of shock heating and radiative and chemical cooling, we consider a non- but nearly isothermal gas with $\gamma < 1.1$, unlike the previous simulations using the isothermal equation of state (EOS) \citep{2012ApJ...747...47T,2014ApJ...790...32W,2014ApJ...782...65S}.
The gas accreting from the protoplanetary disk is initially cool enough to contain molecular hydrogen H$_2$. When and after the cold flows pass through accretion shocks, however, the temperature becomes high enough for dissociation of H$_2$ to occur, which affects the equation of state, as pointed out by \citet{2016MNRAS.460.2853S}. 
In the temperature range where H$_2$ is being dissociated, the effective $\gamma$ becomes approximately 1.1 \citep[e.g.][]{2000ApJ...531..350M}. 
The radiative cooling can further reduce the effective $\gamma$. 
We set up our model based on the above consideration in this paper. We will conduct radiation-hydrodynamic simulations with realistic microscopic physics to give more conclusive results in our future study.
In our fiducial model, $\gamma=1.01$. 
We also perform simulations with different values of $\gamma$, and we show the case with $\gamma=1.05$ for comparison.

\begin{figure}
\epsscale{1.0}
\plotone{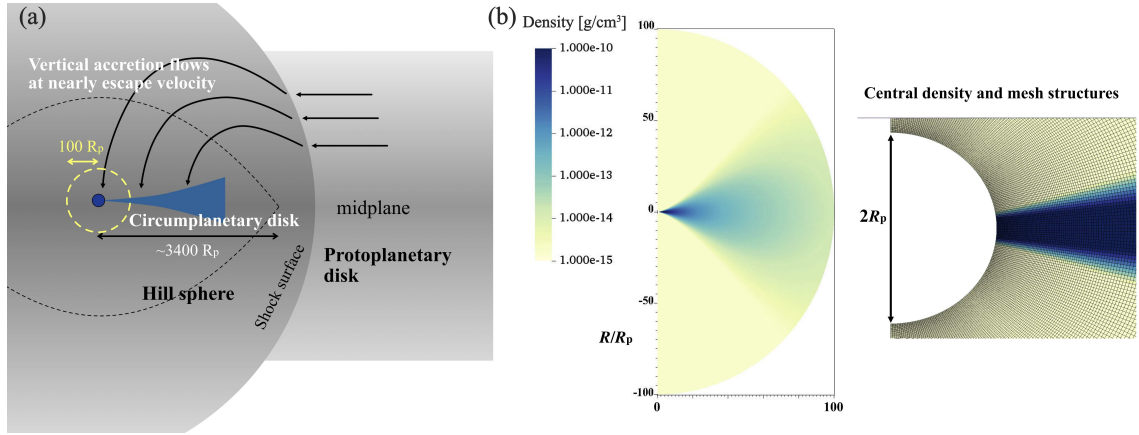}
\caption{(a) Schematic illustration of our protoplanet-CPD model. We solve the accretion process in the region within the yellow dashed circle. (b) The initial density distribution of our 2D model. The left panel shows the entire simulation domain, while the right panel displays the central region just around the proto-giant planet. The mesh structure is also shown in the right panel, where the empty half circle represents the protoplanet. Note that the protoplanetary radius $R_{\rm p}$ is assumed to be $2R_{\rm J}$.
\label{fig:ic_density}}
\end{figure}

Figure~\ref{fig:ic_density}(a) outlines our model. We consider the situation for PDS~70b, where the protoplanet with mass of $\sim 10~M_{\rm J}$ and radius of $\sim 2R_{\rm J}$ is embedded in the protoplanetary disk around the host star with mass of $\sim 0.8~M_{\odot}$. The distance of the protoplanet from the host star is $\sim 20$~au. The Hill radius is $\sim 3.2$~au, which corresponds to $\sim 3.4\times 10^{3}~R_{\rm p}$. The gas is supposed to accrete from the protoplanetary disk onto both CPD and protoplanet \citep{2012ApJ...747...47T}, as indicated by the black arrows; the accretion flows first pass through a weak shock surface and nearly vertically enter the Hill sphere (indicated by dashed lines). In this study, we focus on the accretion dynamics in the region within the yellow dashed circle. The protoplanetary disk is not included in our simulations and is treated as an outer boundary condition, as described below. Under the assumption that the non-axisymmetric processes can be ignored in this central region, we use the 2D spherical polar coordinates $(r,\theta)$ centered on the protoplanet center, where $r$ and $\theta$ are the radius and latitudinal angle measured from the north pole, respectively. The simulation domain is $(R_{\rm p}, 0)\le (r,\theta) \le (100R_{\rm p},\pi)$. In our fiducial model, the protoplanetary mass ($M_{\rm p}$) and radius ($R_{\rm p}$) are $12M_{\rm J}$ and $2R_{\rm J}$, respectively, according to \citet{2019ApJ...885L..29A}. With this normalization, the outer boundary is located approximately at 0.1~au away from the center. 

The calculated domain of our 2D models is resolved with $300 \times 200$ meshes. The mesh spacing is uniform in the $\theta$-direction. The radial mesh size is proportional to the radius, $r$, so that the ratio of the radial to latitudinal mesh sizes is nearly kept constant. The radial and latitudinal mesh sizes just around the protoplanet are approximately $5\times10^{-3}R_{\rm p}$ and $1.6\times 10^{-2}R_{\rm p}$, respectively. We adopt a much finer spatial resolution than previous studies. The finest mesh size in \citet{2012ApJ...747...47T} is $\sim 0.25 R_{\rm J}$ for the model of a Jovian-mass protoplanet placed at $\sim 5$ au from a solar-mass star. 
The protoplanetary scale in \citet{2020arXiv200209918S} is not spatially resolved, and its gravitational potential is artificially softened on a spatial scale of $\sim 40R_{\rm J}$ for the model of a protoplanet with a mass of $10M_{\rm J}$ at 5.2 au from a solar-mass star. Therefore, the accretion heating around the protoplanet is suppressed by the gravitational softening. Their finest spatial resolution is approximately $2R_{\rm J}$, which is insufficient to resolve the stratified gas structure on the protoplanetary scale.
In this study, we combine the high-spatial-resolution global simulations with the local radiation hydrodynamic model of the shock-heated accretion flow, as described in Section~\ref{sec:aoyama-model}.

\subsection{Boundary and Initial Conditions}
We describe our CPD model and the inner and outer boundary conditions. The inner and outer boundaries correspond to the surfaces of the filled blue and dashed yellow circles, respectively, in Figure~\ref{fig:ic_density}(a). The protoplanetary surface is not resolved and is treated as an outgoing inner boundary. The data in the cells touching to the protoplanetary radius are passed to the 1D radiation-hydrodynamic model (see Section~\ref{sec:aoyama-model}).

To construct the outer boundary condition, we referred to the results of the 3D hydrodynamic simulation by \citet{2012ApJ...747...47T}, where the accretion flows almost vertically falls onto the protoplanet and CPD from the protoplanetary disk, almost at the local escape velocity. The mass flux onto the CPD surface is nearly constant with radius, and the specific angular momentum is proportional to $R^{a}$, where $R$ is the cylindrical radius and the index $a$ is approximately 1.0-1.5. We take $a=1.5$ in this study. The spherical polar coordinates $(r,\theta)$ is related to the cylindrical coordinates $(R,z)$ by the relations $R=r\sin{\theta},z=r\cos{\theta}$.
Following their results \citep[see Figure~15 of][]{2012ApJ...747...47T}, at the outer boundary, we set 
\begin{align}
    v_{\rm z}&=-v_{\rm esc}(r_{\rm out}){\rm sign}(z),\\
    v_{\phi} & = b v_{\rm K}(R_p)\left( \frac{r\sin{\theta}}{R_p}\right)^{1/2},
\end{align}
within the $\theta$ range defined by $|\theta-\pi/2| > \pi/4$ ($\equiv \Delta \theta_{\rm BC}$).
Otherwise, no mass is injected, and only the outgoing mass flux is allowed.
Here, $v_{\rm z}$ is the velocity component in the vertical ($z$-) direction, $v_{\rm esc}(r)$ is the escape velocity at the radius of $r$ (= $\sqrt{2GM_p/r}$), $r_{\rm out}$ is the radius of the outer boundary whose value is 100~$R_p$ (= 200 $R_\mathrm{J}$). Also, $v_\phi$ is the azimuthal component of velocity, $v_\mathrm{K}$ is the Keplerian velocity around the protoplanet, and the numerical factor $b$ is assumed to be 0.01, based on \citet{2012ApJ...747...47T}. 
The temperature and density are fixed to those of the protoplanetary disk gas, denoted by $T_0$ and $\rho_0$, respectively. We take $\rho_0 = 1.7\times10^{-15}~{\rm g~cm^{-3}}$ and $T_0=1290~{\rm K}$. 

The injection rate of mass at the outer boundary $\dot{M}_{\rm out}$ is therefore written as 
\begin{align}
    \dot{M}_{\rm out} = -2\pi r_{\rm out}^2 \rho_0 v_{\rm z}(r_{\rm out}) \int^{\pi}_{0} \cos{\theta}\, d(\cos{\theta});
\end{align}
$\dot{M}_{\rm out}$ is almost constant with time, but it slightly fluctuates because of the gas motion mainly within $|\theta-\pi/2| < \Delta \theta_{\rm BC}$.

The CPD model is also based on the result of \citet{2012ApJ...747...47T}, where the midplane density of their isothermal disk is roughly inversely proportional to $r^3$. We construct the uniform-temperature, nearly hydrostatic disk model as follows. The hydrostatic balances in the $r$ and $\theta$ directions are respectively expressed as
\begin{align}
\frac{\partial p}{\partial r} -\frac{\rho v_{\phi}^2}{r}&= -\frac{\rho G M_p}{r^2},\label{eq:hydrostatic-r}\\
\frac{\partial p}{\partial \theta} & = \rho v_{\phi}^2 \frac{\cos{\theta}}{\sin{\theta}}.\label{eq:hydrostatic-theta}
\end{align}
The temperature $T$ is set to $T_0$ in the entire simulation domain, and the density is written as
\begin{align}
    \rho(r,\theta) = \rho_{d0} \left( \frac{R_p}{r}\right)^3 f(r,\theta),
\end{align}
where $\rho_{d0}$ is the midplane density at $r=R_p$, and $f(r,\theta)$ is a function to describe the density distribution out of the midplane ($f(r,\theta=\pi/2)=1$ by definition). We take $\rho_{\rm d0}=1.7\times 10^{-8}~{\rm g~cm^{-3}}$.
The gas pressure $p$ is therefore 
\begin{align}
    p(r,\theta)=\rho R_g T = \rho_{d0} c_{\rm iso}^2 \left( \frac{R_p}{r}\right)^3 f(r,\theta),
\end{align}
where $R_g$ is the gas constant and $c_{\rm iso} = \sqrt{R_g T_0}$ is the isothermal sound speed. In our simulations, we set the temperature $T_0$ so that $c_{\rm iso}^2/v_{\rm K}(R_p)^2=10^{-3}$.
From Equation~(\ref{eq:hydrostatic-r}), we obtain
\begin{align}
    v_{\phi}^2 = v_{\rm K}(r)^2 -3 c_{\rm iso}^2 + c_{\rm iso}^2 \frac{\partial \ln{f(r,\theta)}}{\partial \ln{r}}.\label{eq:vphi2}
\end{align}
Here we note that the deviation of the azimuthal velocity from the Keplerian velocity should be very small around the midplane because our disk is cold ($c_{\rm iso}^2 \ll v_K^2$). From Equation~(\ref{eq:hydrostatic-theta}), we get
\begin{align}
    c_{\rm iso}^2 \frac{\partial \rho}{\partial \theta} = \rho v_{\phi}(r,\theta)^2\frac{\cos{\theta}}{\sin\theta}.
\end{align}
We analytically solve this equation by assuming that the azimuthal velocity is nearly the Keplerian velocity ($v_{\phi}(r,\theta)\approx v_{\phi}(r)=v_{\rm K}(r)$). This assumption holds true especially around the midplane. The result is
\begin{align}
    f(r,\theta)\approx |\sin\theta|^{v_{\phi}(r)^2/c_{\rm iso}^2}.
\end{align}
We check the validity of the above assumption. Using the functional form of $f(r,\theta)$, we get
\begin{align}
    \frac{\partial \ln{f(r,\theta)}}{\partial \ln{r}} = - \frac{v_{\rm K}(r)^2}{c_{\rm iso}^2}\ln{|\sin\theta|} > 0.
\end{align}
Then, Equation~(\ref{eq:vphi2}) becomes
\begin{align}
    v_{\phi}^2 \approx v_{\rm K}(r)^2(1-\ln{|\sin\theta|}) -3 c_{\rm iso}^2.
\end{align}
The factor $(1-\ln{|\sin\theta|})$ is 1.0-1.1 approximately for $65^\circ < \theta < 125^\circ$. Therefore, the assumption holds for a large body of the disk.
The initial density distribution is shown in Figure~\ref{fig:ic_density}(b). The mesh structure around the central region is also shown.

Physical viscosity is ignored in our model, which means that the viscous accretion through the disk midplane does not occur. 
As we will see later, the accretion onto the protoplanet occurs in the disk surface layers and outside the disk.
The above assumption of the inviscid disk does not affect our results related to the accretion shock structures.

\subsection{Modeling of H$\alpha$ emission}\label{sec:aoyama-model}

To quantify the H$\alpha$ emission from the protoplanet--CPD system,
we combine the 1D radiation-hydrodynamic model of the shock-heated accretion flow, which was developed by \citet{2018ApJ...866...84A}, with the 2D high-resolution hydrodynamic model introduced in the previous subsection. 
As with \citet{2018ApJ...866...84A}, we take into account the effects of chemical reactions, excitation/de-excitation of hydrogen atoms, and radiative transfer. We calculate the collisional and radiative transitions between energy levels of hydrogen atoms in a time-dependent way. The gas is assumed to consist of four elements including hydrogen, helium, carbon, and oxygen with the solar abundances. The main input parameters are the preshock velocity $v_0$ and the number density of hydrogen nuclei $n_0$. The jump conditions are analytically applied at the shock under the strong shock assumption.

The model by \citet{2018ApJ...866...84A} predicts that the H$\alpha$ emission in the postshocked region will originate from a very thin layer, the thickness of which is, for instance, $<100~{\rm km}$ for $v_0\sim$100~km~s$^{-1}$ and $n_0 \sim 10^{11}$~cm$^{-3}$. Various microscopic processes including the chemical reactions and excitation/de-excitation of hydrogen atoms proceed on a timescale shorter than a second.
Because it is extremely computationally expensive to simulate the accretion process on an astronomical scale while simultaneously resolving the very thin H$\alpha$-emitting layers, we decide to take the following approach. 
We first perform high-resolution 2D hydrodynamic simulations of the protoplanet--CPD system. We then pass the data to the 1D radiation-hydrodynamic model to calculate the H$\alpha$ emission from the protoplanet--CPD system.
For the calculation of the protoplanetary emission, we measure the physical quantities at the protoplanetary surface. The data are used to model the emission from the accretion shocks, which are unresolved in the 2D simulations.
For the calculation of the CPD emission, we identify the location of the accretion shocks formed above CPD in the 2D simulations and measure the physical quantities in the upstream of the shocks. We use the data in both the northern and southern hemispheres to calculate the luminosity but only use the data in the northern hemisphere for modeling the H$\alpha$ line profile.
This is because the luminosity should be defined as the energy flux integrated in the whole solid angle ($4\pi$), while the observed emission comes only from a hemisphere.

\section{Numerical Results} \label{sec:results}
\subsection{Overview}
\begin{figure}
\epsscale{0.5}
\plotone{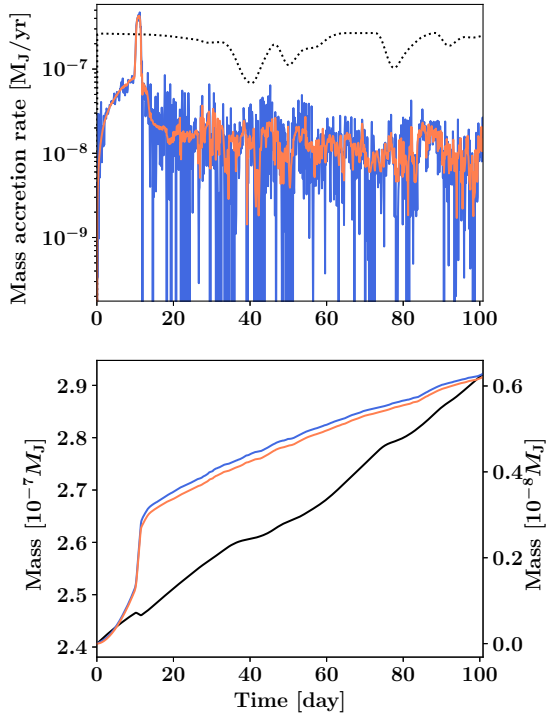}
\caption{The accretion history. Top: the total mass accretion rate measured at the protoplanetary surface is shown by the blue line, and the rate of mass accretion by fast accretion flows is indicated by the red line. The mass injection rate from the outer boundary is shown by the dotted line. Bottom: The black line shows the mass in the simulation domain (between the inner and outer boundaries, left axis). The blue line indicates the total mass supplied by the accretion, while the red line indicates the mass supplied by the fast accretion only (right axis).
\label{fig:accretion-rate}}
\end{figure}

The mass accretion history is shown in Figure~\ref{fig:accretion-rate}. In the top panel, the blue line shows the total mass accretion rate measured at the protoplanetary surface, while the red line denotes the rate of mass accretion only by fast accretion flows whose speed is higher than 30\% of the Keplerian velocity at the planetary surface ($v_{\rm K}(R_{\rm p})\approx 100~{\rm km~s^{-1}}$). The mass injection rate at the outer boundary is shown by the dotted line. 
From the bottom panel of Figure~\ref{fig:accretion-rate}, one can see that the mass is mainly carried to the protoplanet by the fast flows. The accretion rate is $\sim 10^{-8}M_{\rm J}~{\rm yr^{-1}}$ if it is averaged on a 10 day timescale. However, the accretion rate is highly time-variable on shorter timescales; the accretion rate varies by an order of magnitude on a timescale of days or less. The bottom panel shows the temporal evolution of the mass contained in the simulation domain (between the inner and outer boundaries).

We stop our calculations at $t \sim$ 100 days because the CPD mass increases by $\sim$20\% from the initial mass and, thereby, the disk surface density profile largely deviates from the initial setting. The rapid mass increase is due to the lack of angular momentum exchange processes in our models. We do not include the explicit viscosity because possible angular momentum exchange processes in CPD remain unclear, and we have no clear ideas about the typical value for viscosity. For these reasons, we only investigate hydrodynamic processes on timescales shorter than 100 days. Investigation of such short-timescale processes suffices for understanding of the origin of H$\alpha$ emission from an accreting protoplanet system. To study a longer time evolution, we probably need to take into account the mass circulation between the CPD and the protoplanetary disk \citep{2012ApJ...747...47T,2014ApJ...782...65S} or midplane accretion \citep{2002ApJ...580..506T}.

\begin{figure}
\epsscale{1.2}
\plotone{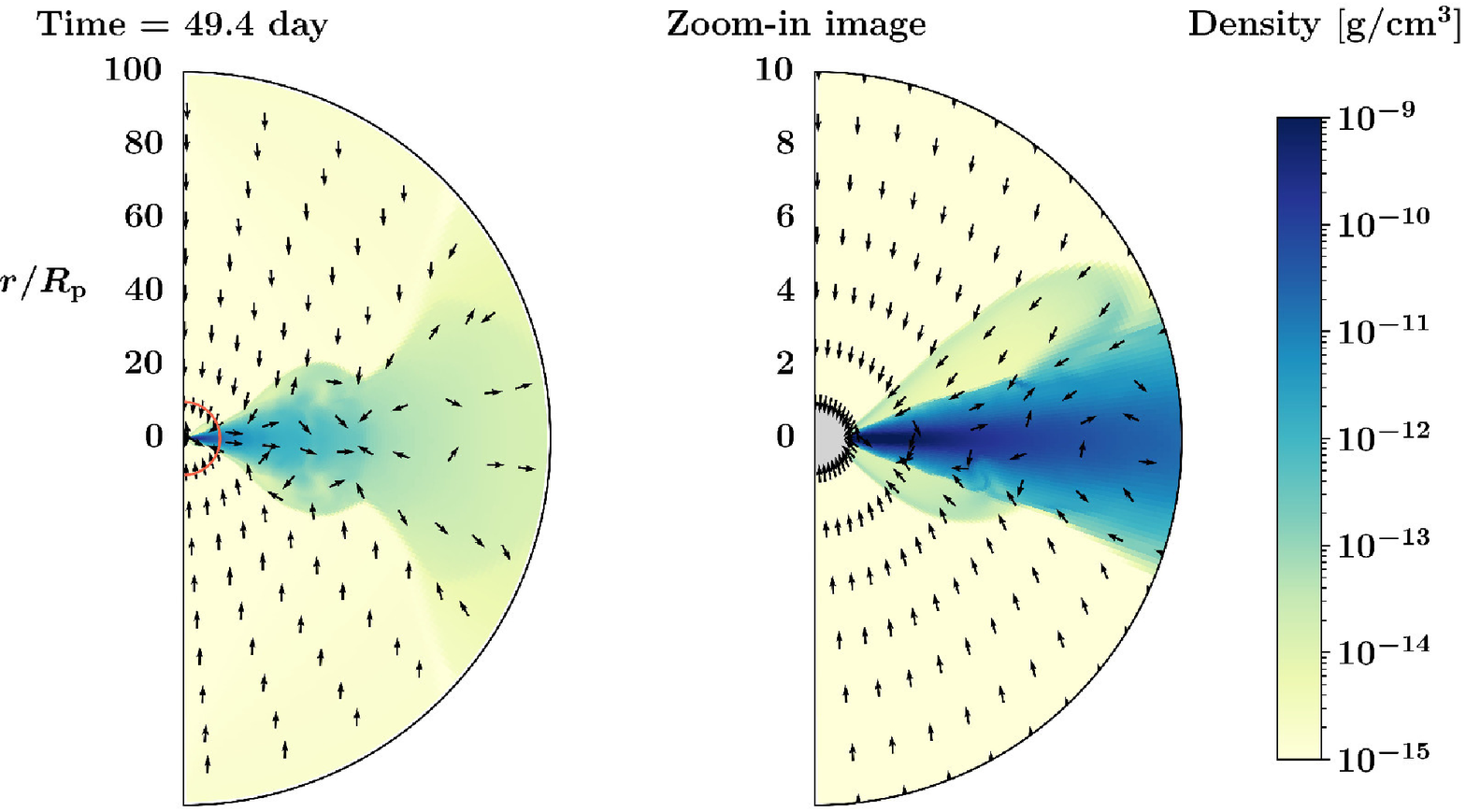}
\caption{Global accretion structure. Left: density map of the entire simulation domain. Arrows indicate the direction of the poloidal velocity (the size does not denote the speed). The length scale is normalized by the protoplanetary radius $R_{\rm p}$. Right: zoom-in image of the central region within $r\le 10 R_{\rm p}$ (indicated by the orange circle in the left panel), where accretion shocks above the CPD can be seen.
\label{fig:accretion-structure}}
\end{figure}

Figure~\ref{fig:accretion-structure} displays the global accretion structure. The left panel shows the entire domain, while the right panel displays the zoom-in image of the central region $r \leq 10 R_p$. Arrows indicate the direction of the poloidal velocity. Note that the size of the arrows does not denote the speed. 
One can discern that the CPD surface structure is largely disturbed by vertical accretion. In the zoom-in image, accretion shocks well above the CPD are clearly seen as the density discontinuities in both the northern and southern hemispheres. Unlike the simulations of \citet{2012ApJ...747...47T} that adopt the isothermal EOS, the accretion shocks are well separated from the CPD surfaces because of the shock heating. Shock heating enables the shocks to propagate toward upstream. The vertical accretion flows change their direction toward the center after passing through the shocks. In other words, the conical shocks converge the accretion flows toward the protoplanet. The convergence occurs because the velocity component normal to the shock is reduced but the parallel component remains unchanged across the shocks (also see $v_{r}$ and $v_\theta$ in Figure~\ref{fig:time-theta}). 
We will later investigate the dynamic properties of the accretion shocks and the CPD surface accretion layers.

Accretion shocks above the CPD surfaces have veen found to vary greatly with time. Such time-variability is caused by complicated radial motions of the accreting gas. As the accreting materials have finite angular momenta whose values depend on their initial radii, they experience a significant centrifugal force around radii where the centrifugal force acting on the gas element balances the gravitational force. The centrifugal force decelerates the CPD surface accretion flows at different radii and at different times, which results in the complex time-variability. 

\begin{figure}
\epsscale{0.5}
\plotone{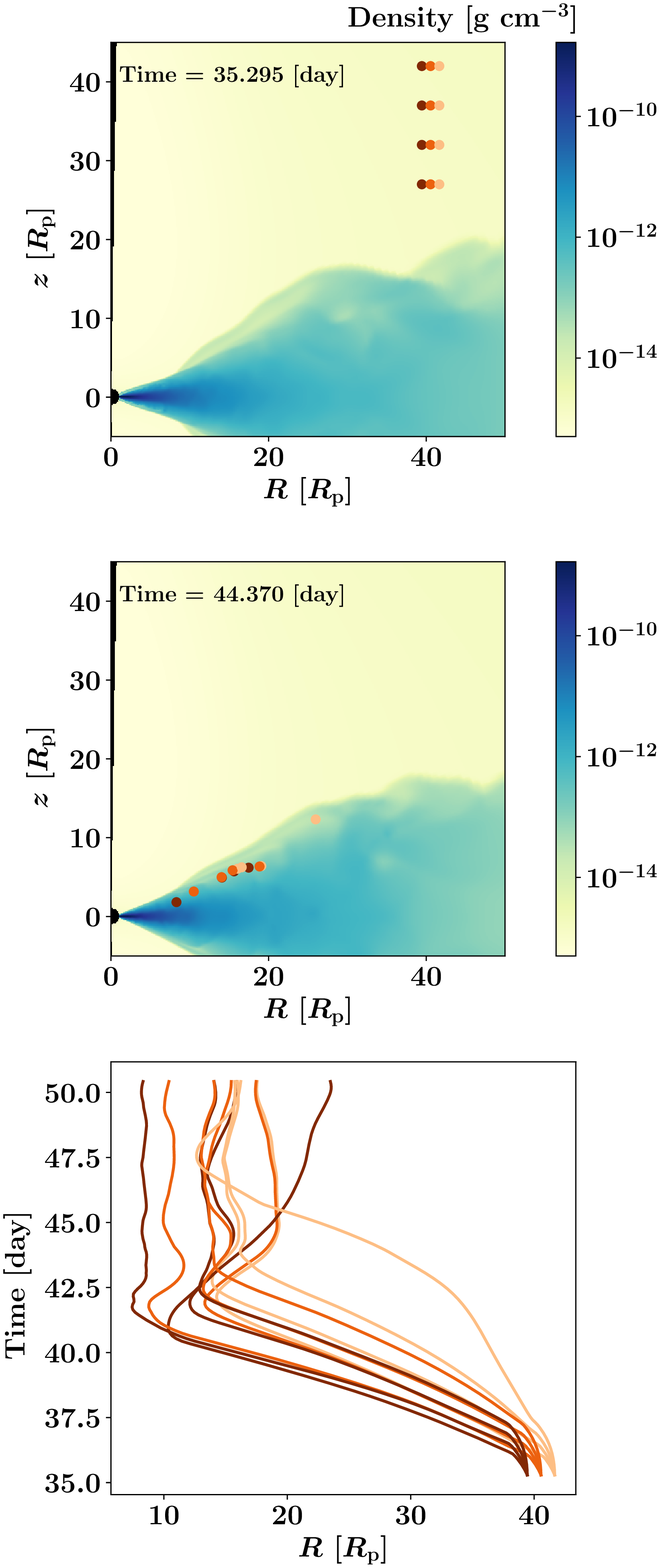}
\caption{Result of the Lagrangian particle analysis. Top and middle panels: the initial and resulting locations of the Lagrangian particles. Bottom: Time--radius diagram of the Lagrangian particles. Each line shows the temporal evolution of the cylindrical radius of a particle. The line colors correspond to the colors of the particles. An animation of the top two panels of this figure is
available. The animation shows the evolution from 35.295 to 50.421 days. The real-time duration of the animation is 12 s.
\label{fig:lagparticle}}
\end{figure}

To investigate the motions of accreting gas in more detail, we perform the Lagrangian particle (test particle) analysis as a postprocessing. Figure~\ref{fig:lagparticle} shows the result. The top panel shows the initial locations of the particles. The background color indicates the density. Those particles are advected in the velocity fields obtained from the hydrodynamic simulation. The middle panel displays the locations of the particles approximately after nine days. In fact, those particles settle and remain around the CPD surface and do not fall onto the protoplanet during the simulation. The bottom panel exhibits the temporal evolution of the cylindrical radius of the particles, where we can see a significant deceleration of particles around at $R\sim 10-15R_{\rm p}$ (the centrifugal radii corresponding to the initial specific angular momenta of the particles are $\sim 10R_{\rm p}$). In addition, some particles show back-and-forth motions on a timescale similar to the local Keplerian periods, which indicates the epicyclic oscillation.

\begin{figure}
\epsscale{0.8}
\plotone{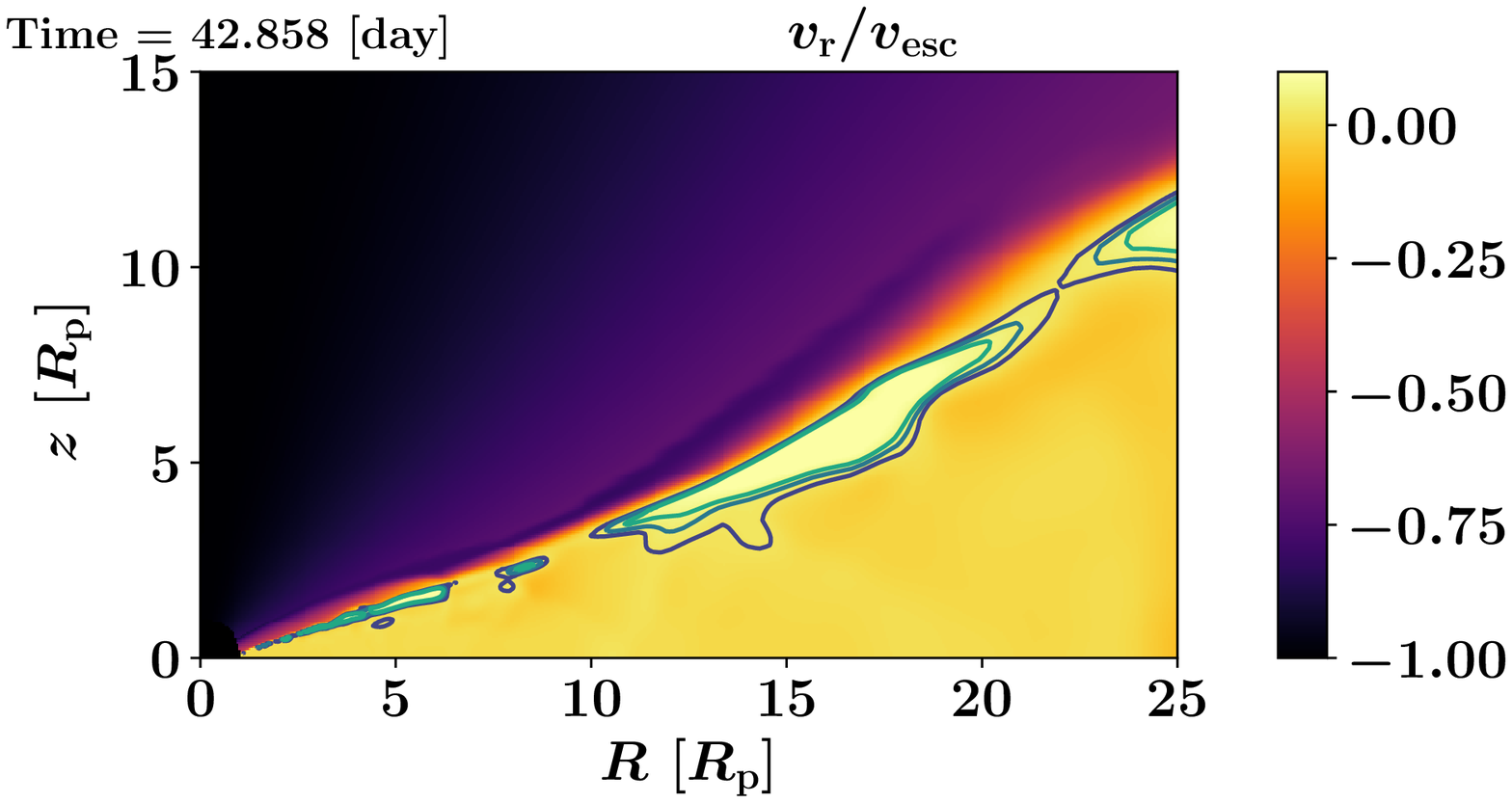}
\caption{Snapshot of the radial velocity normalized by the local escape velocity. The three levels of the contours indicate $v_{\rm r}/v_{\rm esc}=$0.02, 0.04 and 0.06, which highlight the regions moving outward.
\label{fig:vrad2dmap}}
\end{figure}

\begin{figure}
\epsscale{1.2}
\plotone{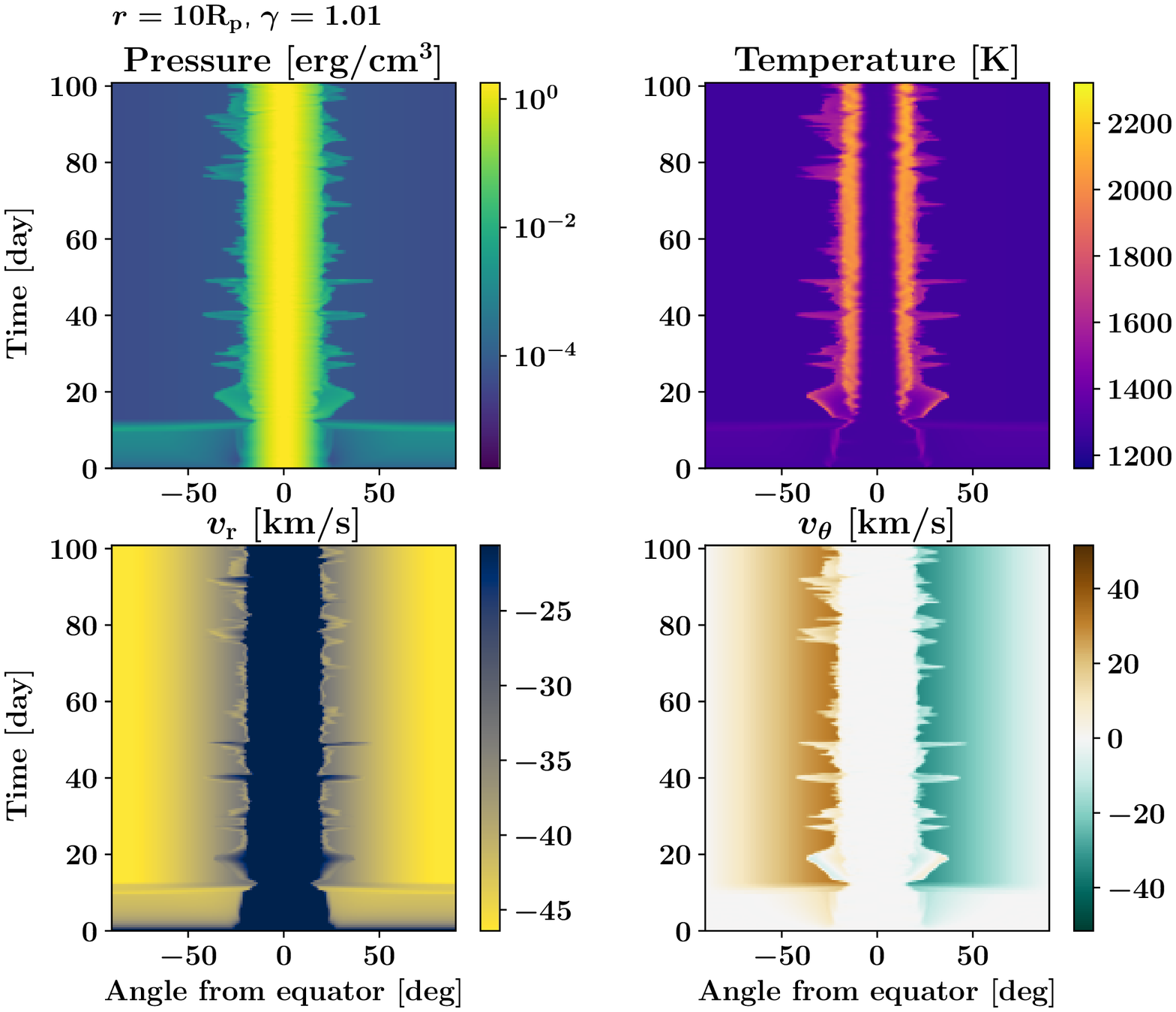}
\caption{Time-variable accretion shocks formed above the disk surfaces. The time-sequenced images of the latitudinal ($\theta$-direction) profiles of four physical quantities measured at the radius of $10R_{\rm p}$.
From the top left to the bottom right, the pressure, temperature, radial component of the velocity, and the latitudinal component of the velocity are shown.
\label{fig:time-theta}}
\end{figure}

Because of the centrifugal force, thin, outgoing flows are formed around the CPD surfaces. Figure~\ref{fig:vrad2dmap} shows a snapshot of the radial velocity normalized by the local escape velocity. The three levels of the contours indicate $v_{\rm r}/v_{\rm esc}=$0.02, 0.04 and 0.06, which highlight the regions moving outward. The outgoing flow pushes back some Lagrangian particles to radii larger than their centrifugal radii. Figures~\ref{fig:lagparticle} and \ref{fig:vrad2dmap} demonstrate the complex radial motions around the CPD surfaces.

Figure~\ref{fig:time-theta} displays the time-sequenced images of the latitudinal profiles of four physical quantities measured at the radius of $10R_{\rm p}$. From the top left to the bottom right, the pressure, the temperature, the radial component of the velocity, and the latitudinal component of the velocity are shown. 
The discontinuities in the pressure profile correspond to the accretion shocks. The shock-heated layers are formed around the CPD surfaces and the shock surfaces are very time-variable. The shock surfaces are sometimes largely elevated from CPD.

\begin{figure}
\epsscale{1.2}
\plotone{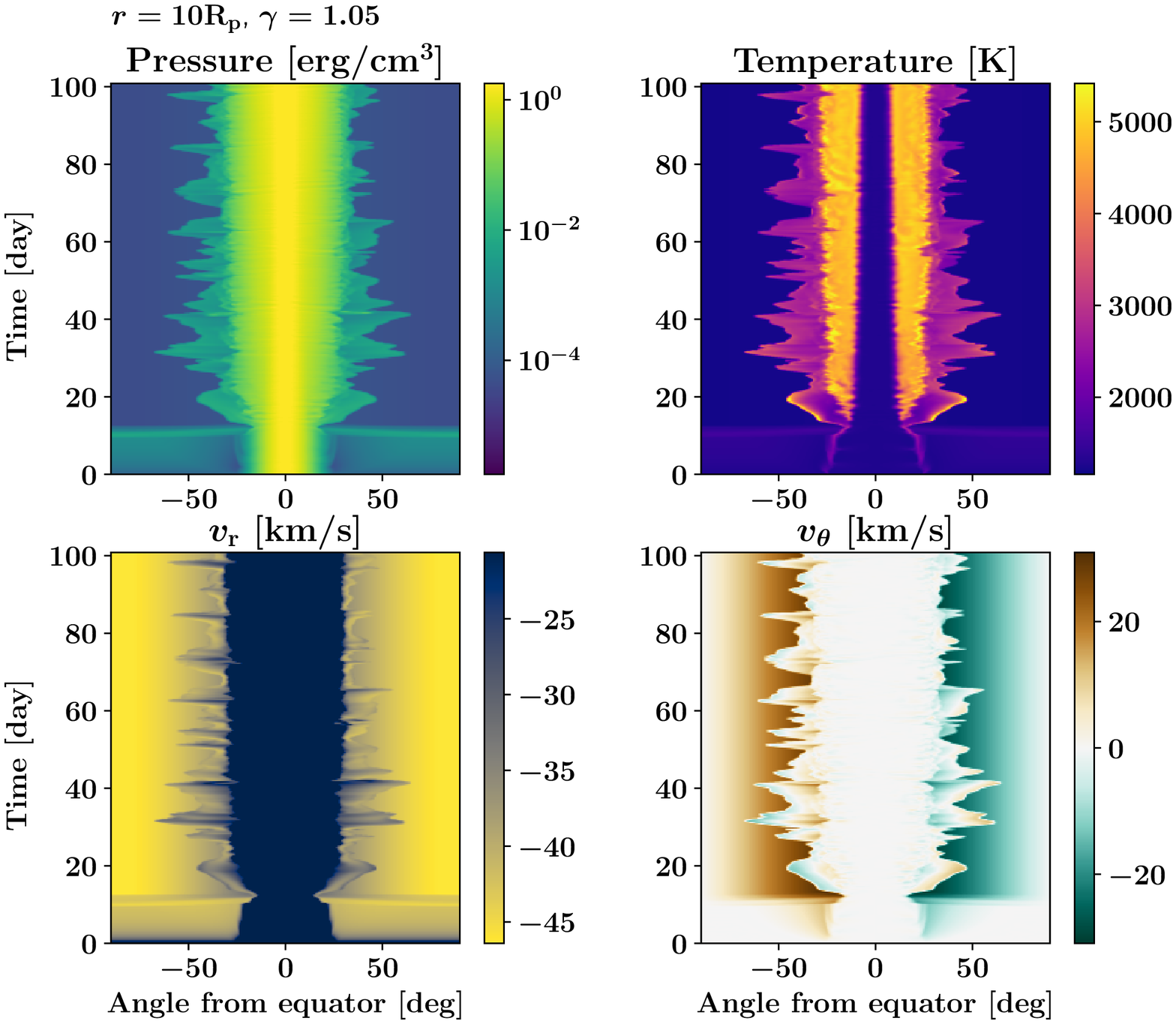}
\caption{Same as Figure~\ref{fig:time-theta}, but for the case of $\gamma=1.05$.
\label{fig:time-theta105}}
\end{figure}

Figure~\ref{fig:time-theta105} is the same as Fig.~\ref{fig:time-theta} but for $\gamma=1.05$ for comparison. The shock surfaces are more elevated and the post-shock regions are thicker than in the case of $\gamma=1.01$. As a larger $\gamma$ leads to a larger increase in temperature, the shocks propagate faster. This is the reason for the higher elevation of the shocks.

The elevated accretion shocks affect both the flow structure and the shock heating and compression.
When the shocks are oblique to streamlines of the upstream flows, only the kinetic energy for the shock-normal velocity component is dissipated.
The higher elevation of the accretion shocks leads to weaker shock heating and compression (and therefore results in the reduction of the H$\alpha$ emission from the CPD surface; we will confirm this point in Section~\ref{subsec:halpha-shock}).
Our model clearly finds the elevation of the shock surfaces because the model spatially resolves the shocked layer and considers the shock heating. No such elevation of the shock surfaces occurs in the previous models that adopt the isothermal EOS, because of the lack of shock heating. Our simulations demonstrate that the shock heating can significantly change the accretion structure around the CPD surfaces.

The accretion rate is modulated mainly through the change in the density of the postshock regions (or disk surface accretion layers; see also Figure~\ref{fig:luminous-area}). The density fluctuation is caused by a combination of some processes. For instance, the radial deceleration due to the centrifugal force leads to a pileup of gas. The dynamical change in the CPD shock angle also modulates the density in the postshock regions by varying the shock strength. The variation amplitude of the accretion rate becomes smaller as $\gamma$ increases because the temperature enhancement across the shock reduces the density enhancement.

\subsection{Accretion onto a Proto-Giant Planet}\label{sec:acc-planet}
\begin{figure}
\epsscale{0.6}
\plotone{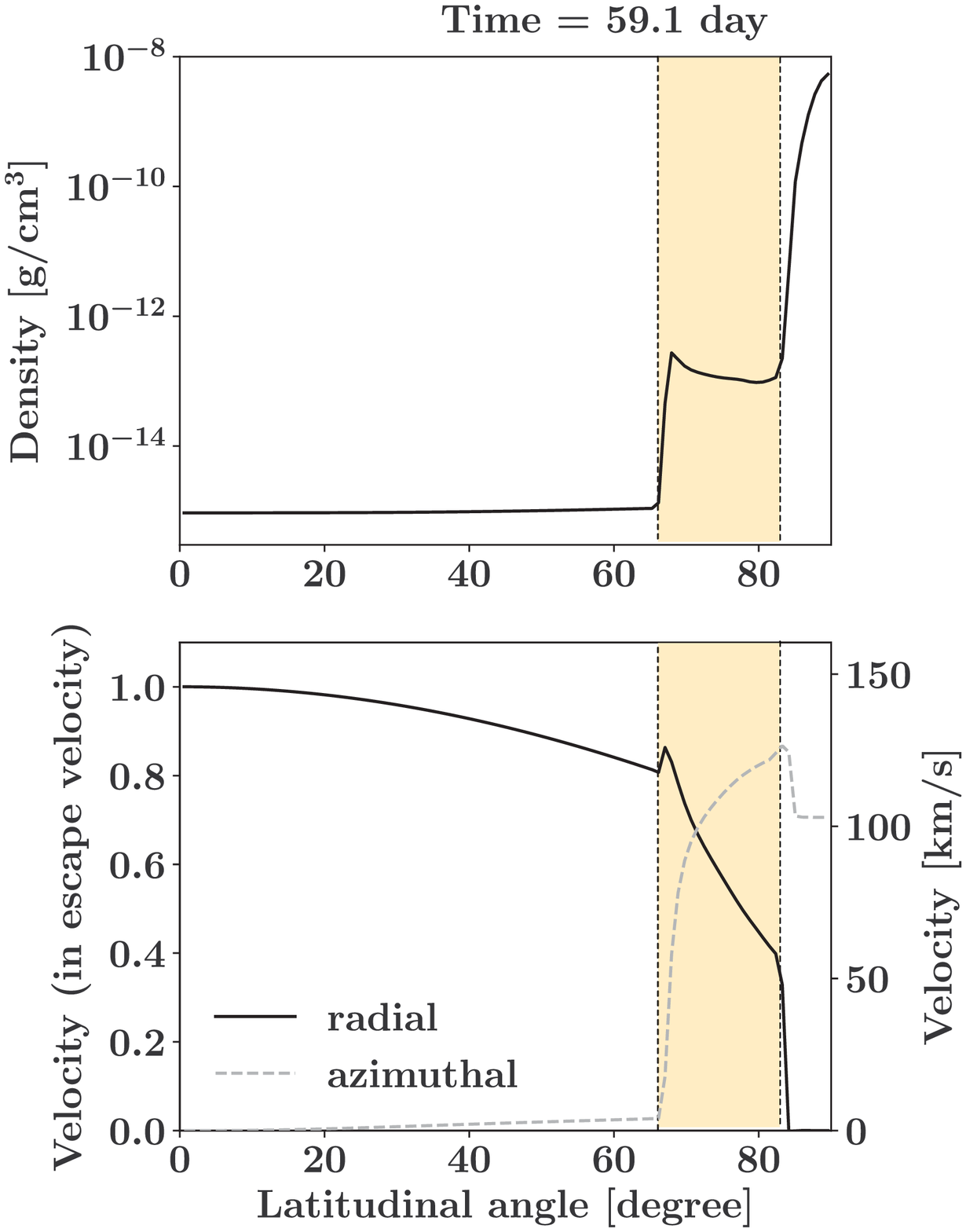}
\caption{The latitudinal ($\theta$-direction) profiles of the density (top) and the velocity (bottom) at the protoplanetary surface. The latitudinal angle is measured from the north pole. In the bottom panel, the solid and dashed lines indicate the radial and azimuthal components of the velocity, respectively. The left axis shows the velocity normalized by the escape velocity at the protoplanetary surface ($\sim 150~{\rm km~s^{-1}}$), while the right axis shows the dimensional velocity. The narrow disk surface layer is indicated by the orange regions. The disk surface layer is located between the tenuous polar region and the dense disk.
\label{fig:accretion-density-velocity-planet}}
\end{figure}

As we will see later, in our model, almost all of the H$\alpha$ emission is produced on the protoplanetary surface. For this reason, we first investigate the accretion region on the protoplanetary surface in detail. Figure~\ref{fig:accretion-density-velocity-planet} shows the density and velocity profiles. The latitudinal angle is measured from the north pole. The density profile can be divided into three regions; the tenuous polar region, the narrow disk surface layer (or the post-CPD-shock region), and the dense disk. The disk surface layer is highlighted in orange. The radial and azimuthal components of the velocity are shown in the bottom panel, where we can also see the three regions. The gas is nearly freefalling onto the polar region, while the radial velocity in the disk surface layer is reduced from the escape velocity by $\sim 20$-60\%. The radial velocity around the disk midplane is negligible in this plot. The azimuthal velocity component is significant both in the disk surface layer and the dense disk, which suggests that the centrifugal force reduces the accretion velocity in the disk surface layer.

\begin{figure}
\epsscale{0.5}
\plotone{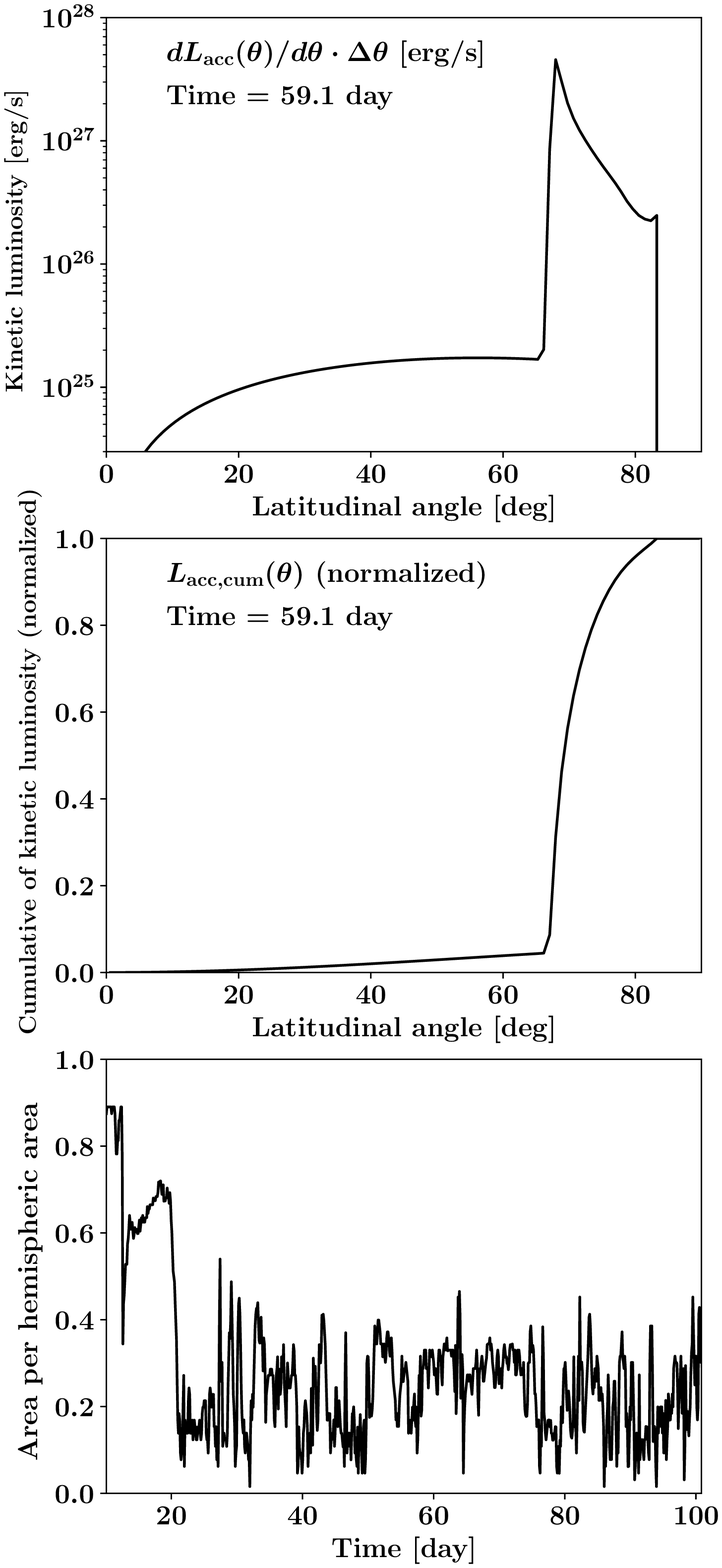}
\caption{The location and temporal evolution of the strongly radiating region on the protoplanetary surface. Top: the latitudinal profile of the accretion (kinetic) luminosity, which is defined by Equation~(\ref{eq:Lacc1}). Middle: the cumulative profile of the accretion luminosity against the latitudinal angle (the angle is measured from the north pole), which is calculated using Equation~(\ref{eq:Lacc2}). Bottom: the temporal evolution of the surface area of the strongly emitting regions in the northern hemisphere. The area is normalized by the hemispherical area, and the vertical axis stands for the filling factor of the accretion region.
\label{fig:emitting-region}}
\end{figure}

The surface area that receives a larger kinetic energy flux should produce stronger radiation. To study the latitudinal distribution of the emission area, we calculate the cumulative sum of the accretion (kinetic) luminosity in the latitudinal direction, $L_{\rm acc,cum}(\theta)$ from the kinetic energy flux $\rho v_{\rm r}^3$: 
\begin{align}
\frac{dL_{\rm acc}(\theta)}{d\theta} &= -2\pi R_{\rm p}^2 \rho(R_{\rm p},\theta)v_{\rm r}(R_{\rm p},\theta)^3 \sin\theta \label{eq:Lacc1}\\
L_{\rm acc,cum}(\theta) &= \int_{0}^{\theta}\frac{dL_{\rm acc}(\theta^\prime)}{d\theta}d\theta^\prime \approx \sum_{\theta^\prime \leq \theta} \frac{dL_{\rm acc}(\theta^\prime)}{d\theta}\Delta \theta
\label{eq:Lacc2}
\end{align}
where $\Delta \theta = \pi/N_\theta$ and the mesh number in the $\theta$ direction and $N_{\theta}$ is 200. Here we ignore the contribution of the azimuthal component of the velocity $v_{\varphi}$ to the kinetic energy flux. The contribution that the protoplanetary surface receives should be defined in the frame of the rotating protoplanetary surface, but the rotational speed of protoplanets is poorly known. For this reason, we only consider the contribution from $v_{\rm r}$.
The top and middle panels of Figure~\ref{fig:emitting-region} show the latitudinal profiles of the kinetic luminosity ($dL_{\rm acc}(\theta)/d\theta$) and the cumulative sum of the kinetic luminosity ($L_{\rm acc,cum}(\theta)$) at $t$ = 59.1~days, respectively, where the kinetic luminosity profile is concentrated approximately between 70$^\circ$ and 80$^\circ$ from the north pole. The narrow region corresponds to the area where the CPD surface accretion flows hit.
This result indicates that the emission caused by accretion mainly originates from the narrow area.
We also note that the narrow area has a substructure where the accretion luminosity takes a sharp peak at the upper edge of the surface accretion layer.

The kinetic luminosity from the polar region is negligible because of the low density. One reason for the low density is that the accretion flows hitting the polar regions do not pass through shocks above CPD and therefore do not experience the shock compression. We note that the maximum shock compression ratio of the gas is $(\gamma+1)/(\gamma-1)=201$ for the gas with $\gamma=1.01$, which is much larger than that for the gas with $\gamma=5/3$. Another reason is related to the outer boundary condition (i.e., the accretion from the protoplanetary disk). The accreting flows with finite angular momenta are vertically falling. Unlike the spherical accretion, their density is not enhanced by the geometrical contraction. The CPD surface accretion flows, on the other hand, have a much higher density because they experienced both the shock compression and the radial contraction.

The bottom panel of Figure~\ref{fig:emitting-region} shows the temporal evolution of the surface area where the CPD surface accretion flows hit. Almost all the protoplanetary H$\alpha$ radiation is indeed produced in this narrow region in our model. In the plot, the emission area is normalized by the hemispheric area. The normalized emission area stands for the filling factor of the accretion region. 
About 15 days after the simulation starts and later, the emission area occupies a few 10\% of the hemisphere on average, but the area highly fluctuates. The fluctuation is caused by the change in the opening angle of the accretion shock formed above the CPD (see Figures~\ref{fig:accretion-structure} and \ref{fig:time-theta}). Comparing Figures~\ref{fig:time-theta} and \ref{fig:emitting-region}, one will notice that the filling factor shows fluctuations on a much shorter timescale than the accretion shocks at $r=10R_{\rm p}$. This indicates that the time-variability of the planetary surface is caused by the superposition of the fluctuations at different radii.

The filling factor of $\mathcal{O}(0.1)$ can be explained as follows. The typical temperature of the disk surface layers, $T_{\rm sl}$, is a few 1,000~K in our fiducial model. The thickness of the layers can be expressed as the pressure scale height, $H(r)=\sqrt{2}c_s/\Omega_K(r)$, where $c_s$ is the sound speed and $\Omega_K(r)$ is the Keplerian angular velocity at a radius of $r$. For the temperature of 2000~K, the ratio of the thickness to the protoplanetary radius is $H(R_p)/R_p \approx 0.06$. The surface accretion layer in a hemisphere will approximately cover the area of $2\pi R_p H(R_p)$. From the above estimates, the filling factor of the disk surface layer in the hemisphere, $f_f$, can be calculated as follows:
\begin{align}
    f_f &\approx \frac{2\pi R_p H(R_p)}{2\pi R_p^2}=\frac{H(R_p)}{R_p}\\
    &\approx 0.06~\left( \frac{T_{\rm sl}}{2\times 10^3~{\rm K}}\right)^{1/2} \left( \frac{R_{\rm p}}{2R_{\rm J}}\right)^{1/2} = \mathcal{O}(0.1).
\end{align}
As shown in the bottom panel of Figure~\ref{fig:emitting-region}, the filling factor takes values close to or larger than this estimated value, except for the initial 20~days.
The time-average filling factor is larger than this value by a factor of a few because the accretion shocks are elevated as a result of shock heating.

\subsection{H$\alpha$ Emission from Shocked Regions}\label{subsec:halpha-shock}
\begin{figure}
\epsscale{1.1}
\plotone{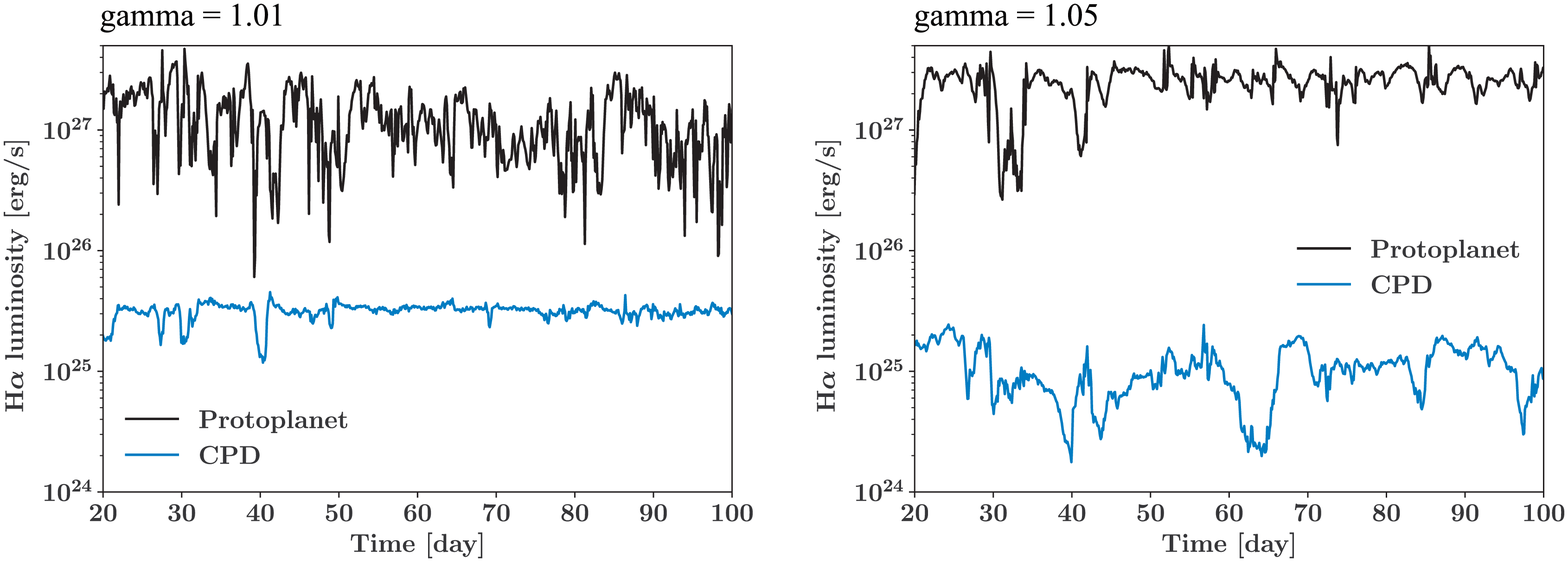}
\caption{The temporal evolution of the H$\alpha$ luminosity. The left and right panels show the results for the cases with $\gamma=1.01$ and 1.05, respectively. The black and blue lines indicate the luminosity from the protoplanet and the circumplanetary disk (CPD), respectively. Here, the emissions in both hemispheres are taken into account.
\label{fig:LHaPlanetDisk}}
\end{figure}

Figure~\ref{fig:LHaPlanetDisk} displays the H$\alpha$ luminosities from the accretion shocks formed on the protoplanetary surface (black) and above the CPD surfaces (blue). Hereafter, the former and the latter are called the protoplanetary luminosity and the disk luminosity, respectively. The figure shows the results of the models with $\gamma=1.01$ (left) and 1.05 (right). We first describe the result of the model with $\gamma=1.01$. The protoplanetary luminosity is approximately one to two orders of magnitude larger than the disk luminosity on average. The protoplanetary luminosity is highly time-variable, as expected from the temporal evolution of the rate of mass accretion by fast accretion flows (Figure~\ref{fig:accretion-rate}), while the disk luminosity is less variable. The disk luminosity remains minor despite its large emission area ($r<10 R_{\rm p}$) because the velocity normal to the shock and the upstream density is smaller than that for the accretion shock on the protoplanetary surface.
We note that the elevation of the accretion shocks due to shock heating also leads to a reduction in the luminosity.

We compare the model with $\gamma=1.05$ to the model with $\gamma=1.01$. The general trend of the model with a larger $\gamma$ is the same as that of the model with a smaller $\gamma$. The protoplanetary luminosity is nearly the same in magnitude between the two models, as the accretion rates are similar. However, the model with the larger $\gamma$ shows a weaker time-variability in protoplanetary luminosity. We note that the gas with the larger $\gamma$ is stiffer (meaning that the density is less sensitive to change in pressure) and shows smaller density fluctuations.
As the modulation in the accretion rate is mainly caused by the density fluctuation in the accretion streams, the time-variability in the protoplanetary luminosity is weaker in the model with the larger $\gamma$.
By contrast, the disk luminosity is smaller than that in the case with the smaller $\gamma$, because the CPD accretion shocks become more vertical to the equatorial plane for models with larger $\gamma$ (see Figure~\ref{fig:time-theta105}). Although the disk luminosity is minor, its time-variability becomes more prominent, because the shock-opening angle changes more largely with time in the model with the larger $\gamma$.

\begin{figure}
\epsscale{1.0}
\plotone{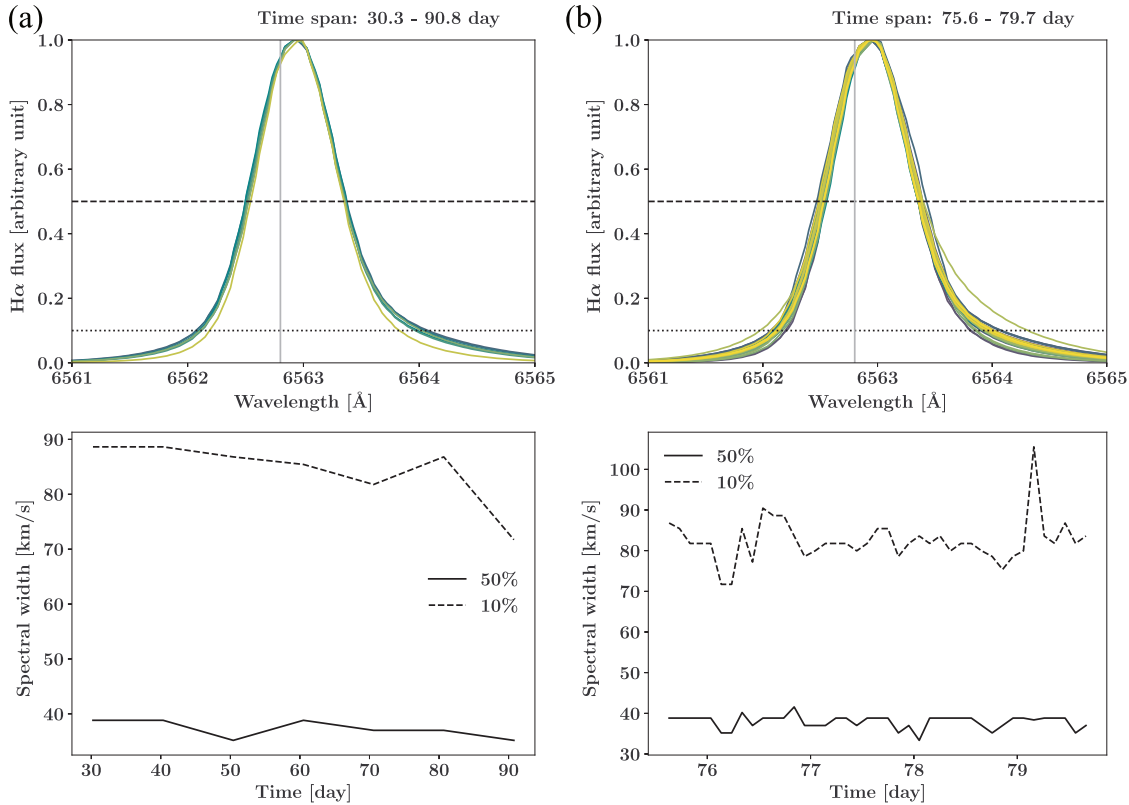}
\caption{The temporal evolution of the spectral profile of the H$\alpha$ line for the model with $\gamma=1.01$. Panel (a) displays the $\sim$10-day cadence result, while panel (b) shows the $\sim$0.1-day cadence result. Note the difference in the sampling time span (indicated at the top of each panel). Top: H$\alpha$ line profiles at different times. The darkest-colored lines show the H$\alpha$ line profile at the beginning of the time spans, and the time increases from darker- to lighter-colored lines with the constant time difference. The vertical gray line shows the H$\alpha$ wavelength in the air (6562.801\AA). The horizontal dashed and dotted lines indicate the 50\% and 10\% full line widths, respectively. Bottom: temporal evolution of 50\% (solid) and 10\% (dashed) line widths in units of km~s$^{-1}$ (the spectral broadening).
\label{fig:HaSED}}
\end{figure}

As the protoplanetary luminosity is dominant in our model, the observed H$\alpha$ line profile is determined only by the protoplanetary emission. 
We focus on the model with $\gamma=1.01$. Figure~\ref{fig:HaSED} displays the H$\alpha$ line profiles from the protoplanet at different times and the measured spectral width. We have performed two different data sampling; panels~(a) and (b) show the $\sim$10-day and $\sim$0.1-day cadence results, respectively. Panel~(a) shows the general behavior on the timescale of several 10~days, while panel~(b) shows more frequent sampling within a duration during 75.6-79.7~days. 
The top panels show the normalized H$\alpha$ line profiles at different times. The darkest-colored lines show the profile at the beginning of the time spans, and the time proceeds from darker- to lighter-colored lines with a constant time difference. 

In the top panel of Figure~\ref{fig:HaSED}~(a),
the line profile remains almost unchanged during the time span. The line center is red-shifted approximately by $6~{\rm km~s^{-1}}$.
The bottom panel shows the temporal evolution of the spectral width. We plot the 50\% (solid) and 10\% (dashed) line widths. The 50\% and 10\% line widths are $\sim$30-40~${\rm km~s^{-1}}$ and $\sim$70-90~${\rm km~s^{-1}}$, respectively. Considering the escape velocity of the protoplanet ($\sim 150~{\rm km~s^{-1}}$), the spectral broadening of the 10\% line width is several tens of percent smaller than the escape velocity. The accretion velocity is reduced by the centrifugal force, as mentioned before (see also Figure~\ref{fig:accretion-density-velocity-planet}). 
The high-cadence data plotted in Figure~\ref{fig:HaSED}~(b) also show a weak time-variability. 
We find a significant line broadening at 79.16~days. As explained in Appendix~\ref{sec:appendix-a}, the large spectral broadening happens when an accretion flow with a small density ($\sim 2\times 10^{10}~{\rm cm^{-3}}$) falls onto the protostar. Regarding observations toward actual objects, the accretion of lower-density flows will lead to a smaller luminosity, which may prevent us from observing such a line broadening. However, we summarize the density dependence of the spectral broadening in Appendix~\ref{sec:appendix-a} for the sake of completeness.

In contrast to the H$\alpha$ luminosity, the line profiles show a much weaker time-variability, which implies the weak time-variability in the major emission region.
We note that the accretion (kinetic) luminosity has a sharp peak at the upper edge of the CPD surface accretion layer (the top panel of Figure~\ref{fig:emitting-region}). Indeed, the global H$\alpha$ line profile reflects the physical condition at this latitude. 

\begin{figure}
\epsscale{0.5}
\plotone{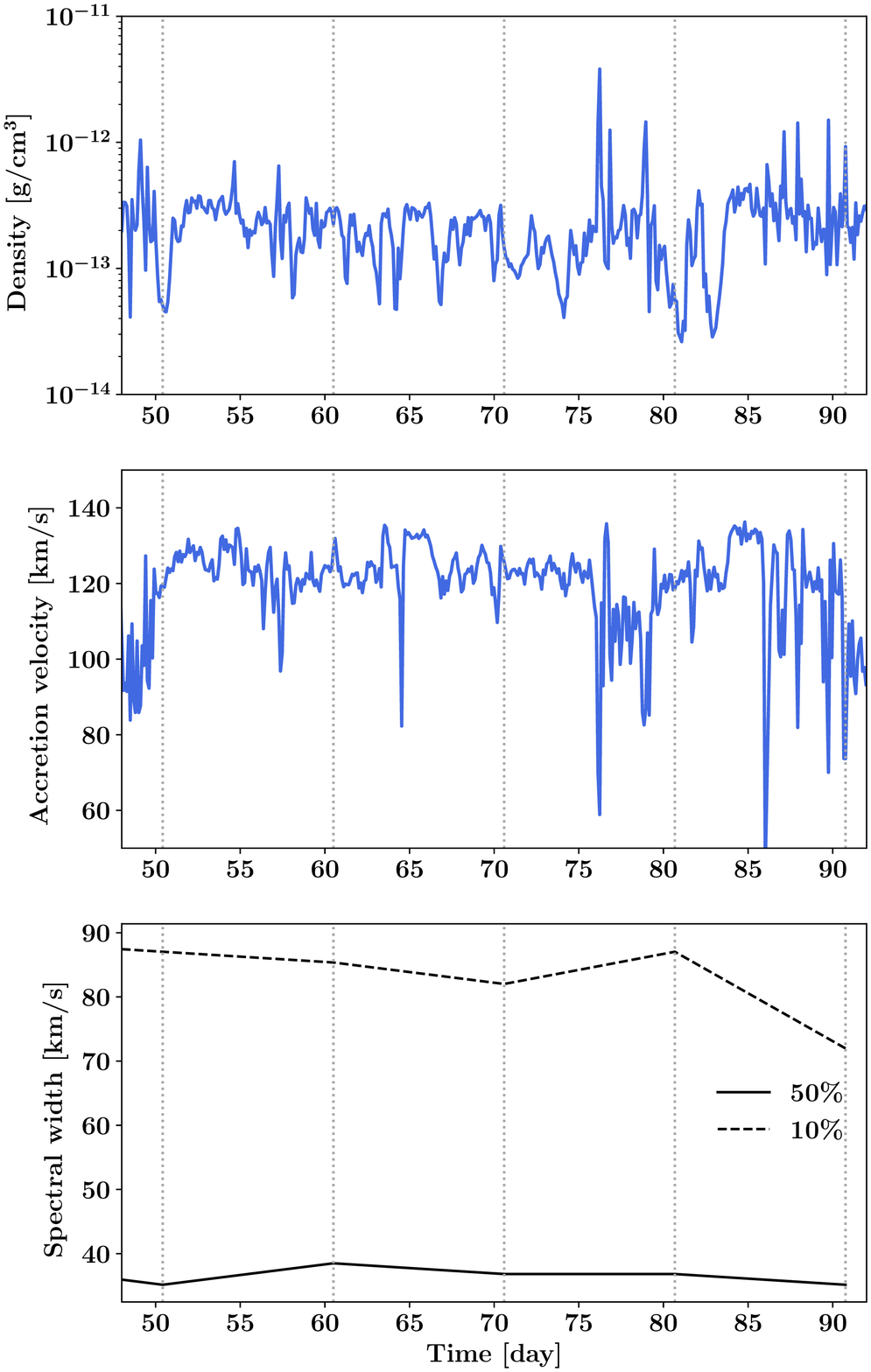}
\caption{The temporal evolution of the density and accretion velocity at the latitude where the accretion (kinetic) luminosity has its maximum. The top panel shows the density, the middle panel shows the accretion velocity, and the bottom panel displays the 50\% (solid) and 10\% H$\alpha$ full line widths (same as in Figure~\ref{fig:HaSED}). The vertical dotted lines indicate the times when the line widths are measured in the period shown here.
\label{fig:luminous-area}}
\end{figure}

Considering that the line width is a function of the accretion velocity and density, we measure them at this latitude.
Figure~\ref{fig:luminous-area} displays the result where the density changes with time by an order of magnitude but the accretion velocity is almost constant except for some short periods. 
Although the density fluctuation has a large amplitude, the density in this range is not sufficiently high to significantly affect the line width through absorption (see Appendix~\ref{sec:appendix-a}). For this reason, the line profile mainly depends on the accretion velocity. 
The accretion velocity at the upper edge of the CPD surface layer remains nearly constant with time because the velocity is determined by the shock jump condition based on the quasi-steady upstream condition.
Therefore, the H$\alpha$ line width remains nearly constant with time. The smaller line width is observed when the accretion velocity becomes occasionally smaller (see the data around $t=90$~days) due to sudden changes in the shock-opening angle.

\section{Summary and Discussion} \label{sec:discussion}
The property of the H$\alpha$ radiation from shocked regions is determined not only by the microscopic physics but also by the accretion structure on the CPD scale.
In this study, we connected a huge scale gap between the micro and macro scales by combining high-spatial-resolution global simulations and local radiation-hydrodynamic simulations. With this approach, for the first time, we directly investigated the origin of H$\alpha$ emission in the vicinity of the protoplanet.
We found that the H$\alpha$ emission is mainly produced in spatially localized areas on the protoplanetary surface. The contribution of the accretion shocks above CPD is approximately one to two orders of magnitude smaller in luminosity. Although the CPD contribution to the H$\alpha$ luminosity is minor, we found that the accretion dynamics around the CPD highly affects the protoplanetary accretion. In our model, the accretion rate onto the protoplanet is $\sim 10^{-8}~M_\odot~{\rm yr^{-1}}$  if averaged on a 10 day timescale. However, the accretion rate is highly time-variable on shorter timescales.

\begin{figure}
\epsscale{1.0}
\plotone{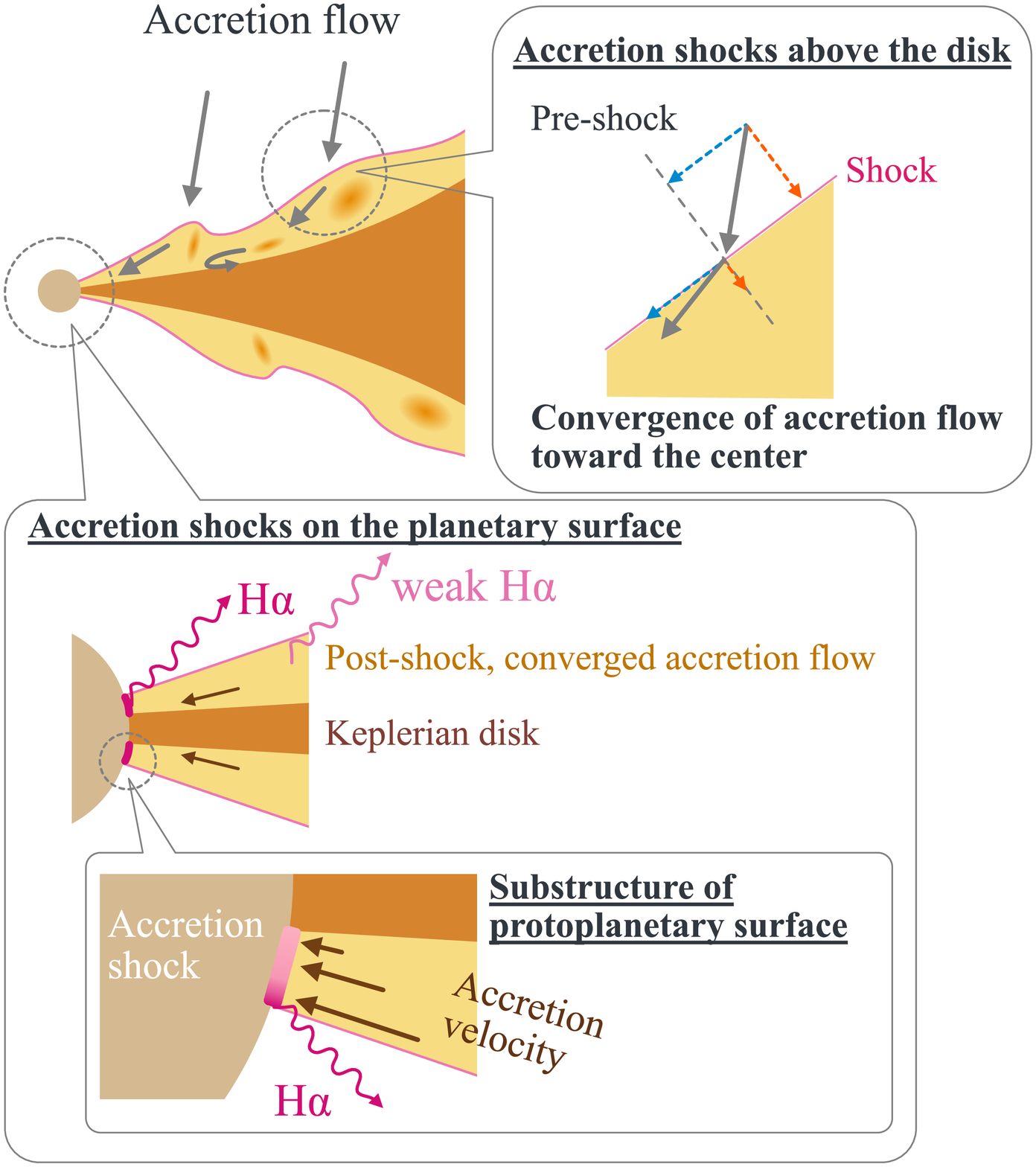}
\caption{Schematic illustration of the accretion structure around the proto-giant planet that we have newly found in this study. 
\label{fig:schematic-diagram}}
\end{figure}

Figure~\ref{fig:schematic-diagram} shows a schematic illustration that summarizes our findings. Our picture can be regarded as an update of that from \citet{2012ApJ...747...47T}, but with a particular focus on the dynamics just around the protoplanet. We studied how the shock heating affects the property of accretion shocks relevant to the observed H$\alpha$ emission. 
Our high-resolution global 2D simulations show that the accretion shocks above the CPD are highly time-variable, although the mass injection from the outer boundary is constant. 
The CPD shock surfaces elevated by the shock heating are found to play two important roles.
We found that the CPD shock surfaces fluctuate due to the centrifugal force and introduce the density fluctuation in the CPD surface accretion layers, which causes the time-variable accretion onto the protoplanet. The centrifugal force decelerates the CPD surface accretion and can drive outward-moving flows (Figures~\ref{fig:lagparticle} and \ref{fig:vrad2dmap}).
Another role is that the elevated CPD shocks converge the vertically accreting flows toward the center without significant heating and compression at the shocks, which suppresses the H$\alpha$ emission from the CPD shocks. The shock heating makes CPD shock surfaces more vertical than in the models that adopt the isothermal EOS. Indeed, a comparison between the two models with different specific heat ratios $\gamma$ shows that the H$\alpha$ emission becomes weaker in the model with a larger $\gamma$ (Figure~\ref{fig:LHaPlanetDisk}).

As shown at the bottom of Figure~\ref{fig:schematic-diagram}, most of the H$\alpha$ emission is produced at the protoplanetary surface (see also Figure~\ref{fig:LHaPlanetDisk}). The accretion rate and the filling factor of accretion areas significantly vary with time (Figures~\ref{fig:accretion-rate} and \ref{fig:emitting-region}, respectively), because of the time-variability in the CPD accretion shocks. However, the time-variability in the H$\alpha$ line profile is much weaker (Figure~\ref{fig:HaSED}). The accretion shocks on the protoplanetary surface have a substructure where the accreting gas flowing around the upper edges emits the H$\alpha$ radiation most strongly. The H$\alpha$ line profile is almost unchanged with time, as the accretion speed at the upper edges shows a weak variability (Figure~\ref{fig:luminous-area}).

\subsection{Model Limitations and Estimation of the Impact of Radiative Cooling}
The results presented in this paper are based on a specific model of accretion from the protoplanetary disk \citep{2012ApJ...747...47T}. If the actual accretion structure differs from our assumption, we should revise the boundary condition and the CPD model accordingly. Nevertheless, we demonstrated that our approach provides a method to overcome the scale gap problem; we can connect the physical processes on the CPD scale and shock-originated emissions on a microscale. The large-scale accretion structure can differ depending on the importance of magnetic fields and radiation \citep{2016MNRAS.460.2853S,2013ApJ...779...59G}. Detailed consideration of the large-scale structure is important to improve our model.

The properties of the CPD surface accretion such as the time-variability and the inclination angle of CPD shocks depend on the thermodynamic processes. In our 2D hydrodynamic models that do not explicitly solve the radiative cooling, the temperature in the CPD postshock region is a few $10^3$~K (Figures~\ref{fig:time-theta} and \ref{fig:time-theta105}). Here we discuss the expected temperature range and the cooling timescale of the postshock region by considering the effect of radiative cooling. We also investigate the impact of the radiative cooling on the property of CPD shocks. 
For this aim, we study the cooling process of the CPD postshock region using a modified version of the \citet{2018ApJ...866...84A} code. We switch off the functions that calculate the energy levels of atoms and molecules because the functions are numerically expensive and prevent us from simulating the cooling process on a timescale longer than $\sim $1-10~s with realistic numerical resources. For the hydrogen line cooling, only the contribution from Ly$\alpha$ is considered under the assumption of the thermodynamic equilibrium between the ground state and the first excited state. For the functional forms of the H Ly$\alpha$ and CO cooling rates, see Appendix~B of \citet{2001Icar..153..430I}.
In contrast to \citet{2018ApJ...866...84A}, this modified model mainly considers the evolution of postshocked gas after the hydrogen line cooling becomes unimportant \citep[see also][for details about the chemical reactions]{2001Icar..153..430I}.

We compare the cooling timescale with the dynamical timescale of the accretion flows. As a reference, we consider the preshock condition of $n_{0}=10^{9}~{\rm cm^{-3}}$. The details about the cooling process are given in Appendix~\ref{sec:appendix-b}. The cooling timescale is found to be $\sim 10^7$~s (Figure~\ref{fig:radiative_cooling}). The cooling is mediated by the H line cooling and CO vibrational cooling. If we take the typical dynamical timescale at the radius $r$ as $t_{\rm dyn}=r/v_{\rm esc}(r)$, 
\begin{align}
  t_{\rm dyn}(r)\approx 10^6 \left( \frac{r}{100R_{\rm p}}\right)^{3/2} ~{\rm s},
\end{align}
which suggests that the dynamical timescale within $r\approx 100R_{\rm p}$ is shorter than the cooling timescale. The dynamical timescale can be comparable to the cooling timescale when $n_0=10^{11}~{\rm cm^{-3}}$. Therefore, in the density range relevant to this study, the temperature of the gas in the post-CPD-shock region will remain several or a few 1000~K during the dynamical timescale.
From this estimate, we consider that the CPD shock dynamics will be affected by the shock heating as shown in this study.

\subsection{Implications for Observational Studies toward PDS~70b}
The accretion rate and the H$\alpha$ luminosity in our fiducial model are $\sim 10^{-8}~M_{\rm J}~{\rm yr}^{-1}$ and $\sim 10^{27}~{\rm erg~s^{-1}}$, respectively, which is consistent with the previous estimate for PDS~70b by different authors \citep{2019NatAs...3..749H,2019ApJ...885...94T,2019ApJ...885L..29A}.
Our model predicts that the H$\alpha$ emission is mainly produced on the protoplanetary surface. 
This means that the line profile is determined solely by the protoplanetary emission. Therefore, H$\alpha$ observations will enable us to directly probe the accretion process in the vicinity of protoplanets.
We note that in our model the accretion through the CPD midplane is assumed not to occur. If the midplane accretion that does not produce accretion shocks is taking place in actual systems, the values of the accretion rate will be underestimated from H$\alpha$ observations.

In our fiducial model, the 10\% full width is approximately 100~km~s$^{-1}$, which is $\sim$30\% smaller than the escape velocity ($\sim 150$~km~s$^{-1}$). This line width is smaller than the value reported by observations with MUSE \citep{2019NatAs...3..749H} ($\sim$220~km~s$^{-1}$). However, as pointed out by \citet{2019ApJ...885...94T}, the spectral resolution of MUSE is $\sim$120~km~s$^{-1}$ around the H$\alpha$ line center, which suggests that the instrumental broadening is non-negligible for the width of $220$~km~s$^{-1}$. Therefore, we consider that the observationally estimated line width gives the upper limit. Higher-spectral-resolution observations are required to reliably resolve the line profile. The line width of $\sim 100~{\rm km~s^{-1}}$ will be difficult to realize if the accretion shocks are formed away from the protoplanetary surface \citep[e.g.][]{2017MNRAS.465L..64S} because the line broadening by the thermal Doppler effect cannot be expected in regions with a shallow gravitational potential.

Our hydrodynamic model predicts that the line width will not largely change with time even if the H$\alpha$ luminosity significantly varies. Monitoring, high-spectral-resolution spectroscopic observations will reveal the property of the accretion dynamics. There is an observational suggestion about the time-variable accretion in PDS 70b; MagAO and MUSE observations at different epochs give different H$\alpha$ luminosities \citep{2018ApJ...863L...8W,2019NatAs...3..749H,2020AJ....159..222H}. Although a direct comparison between different instruments is not straightforward, continuous observations toward PDS 70b will confirm if the accretion process is indeed variable.

Multiple line observations can give a constraint on the extinction. \citet{2020AJ....159..222H} investigated H$\alpha$ and H$\beta$ emissions from PDS 70b, and detected no H$\beta$ emissions. Combining the observational upper limit for the flux ratio $F_{\rm H\beta}/F_{\rm H\alpha}$ with the theoretical model by \citet{2019ApJ...885L..29A}, they estimated the extinction for H$\alpha$ to be $>2$~mag. They discussed that the extinction could be caused by submicron size grains coupled with the gas in the dust gap in the protoplanetary disk. Our hydrodynamic model is consistent with the picture, as most of the H$\alpha$ emissions are produced at the upper edge of the protoplanetary shocked areas and the emissions are not veiled by dense CPD gas. To draw a more reliable conclusion, the dust distribution in the dust gap should be investigated in more detail.

\subsection{Comparison of Our Hydrodynamic Model with the Magnetospheric Model}
The magnetospheric accretion scenario for accreting young stars has commonly been applied to accreting proto-giant planets \citep{2018AJ....155..178B,2019ApJ...885...94T}, mainly to account for the production of hot, H$\alpha$-emitting gas and the spin-down of protoplanets. Here, magnetospheric accretion denotes a type of accretion mode where a strong magnetic field of the central object (protoplanet) decelerates the rotating disk gas to drive a freefall, supersonic accretion along the magnetic field line \citep[for magnetospheric accretion in classical T Tauri stars, see a review by][]{2016ARA&A..54..135H}. The emission regions are spatially localized on the protoplanetary surface, and the filling factor of the accretion areas is expected to be much smaller than unity.
In this study, we showed that the accreting gas onto a proto-giant planet can produce a sufficient H$\alpha$ luminosity and a large H$\alpha$ line width of $\sim 100$~km~s$^{-1}$ even in the absence of the protoplanetary magnetosphere. 
Here we compare our hydrodynamic model and the magnetospheric model.

In the magnetospheric model, the accreting gas is assumed to fall freely onto the protoplanet from a few to several protoplanetary radii, as a result of the rapid angular momentum loss by the strong protoplanetary magnetic fields \citep[for classical T Tauri stars, see][]{1998ApJ...509..802C}. Therefore, the accretion speed is similar to the escape velocity. In classical T Tauri stars, the accretion areas are seen as hot spots. The emissions from accretion shocks often show modulation due to stellar rotation, which suggests the localization of accretion shocks \citep{2007prpl.conf..479B}. The filling factor of the accretion area is estimated to be typically $\mathcal{O}(10^{-3})$-$\mathcal{O}(10^{-2})$ \citep{2004AJ....128.1294C,2013ApJ...767..112I}.
In our hydrodynamic model, the filling factor is $\mathcal{O}(0.1)$ (see Section~\ref{sec:acc-planet}), although the value will depend on the cooling process through the change in the opening angle of CPD shocks. The accretion area will be seen as a hot ring, rather than a localized hot spot. Therefore, the modulation of the emission profile due to protoplanetary rotation will be insignificant. The accretion velocity at the most H$\alpha$-luminous region is $\sim$20\% smaller than the escape velocity because of the centrifugal force. The accretion velocity is reduced by the centrifugal force as the rapid angular momentum loss of accreting matters is absent in the hydrodynamic model. This result indicates that if the hydrodynamic accretion occurs, we will underestimate the protoplanetary mass by several tens of percent if we estimate it by assuming that the accretion speed is the freefall velocity (the escape velocity). 
Regarding the line profile, the magnetospheric accretion will produce a broader line width than the hydrodynamic model for a protoplanet with the same mass, because both the accretion velocity and density are larger due to the smaller filling factor. To investigate the difference in the line profiles, spectroscopic observations with higher spectral resolution than the current MUSE will be required.

To distinguish the two accretion modes, both high-time-cadence observations and multiple line observations with a high spectral resolution will be necessary. An important way will be looking for the periodic modulation caused by the protoplanetary rotation that will appear in the magnetospheric accretion case. The detection of a periodic modulation is a strong indication of magnetic accretion, although independent spectroscopic measurements of the protoplanetary rotation may be required to draw a more robust conclusion. If the periodic time-variability is not observed, we need to examine the possibilities of both modes. Even in the magnetospheric accretion, the modulation can be caused by the CPD shock fluctuation, because the CPD shock will extend to a much larger scale than the magnetospheric size. If the CPD surface accretion dominates the midplane accretion in mass accretion rate, the CPD shock fluctuation will cause the time-variability as described in this study. To identify the accretion modes in the case of nonperiodic modulations, it will be useful to estimate the extinction of the emission from the shocked regions using multiple line observations \citep{2020AJ....159..222H}. In the magnetospheric accretion mode, a time-variable extinction by accretion columns is expected \citep{Bouvier_1999,Alencar_2010}, while such extinction will be absent in the hydrodynamic accretion mode. Therefore, a combination of high-time-cadence observations and multiple line observations will give constraints on the accretion mode.

Without the effect of protoplanetary magnetic fields, the angular momentum of the protoplanet will continuously increase because of the absence of efficient angular-momentum-loss mechanisms. Because observations toward the planetary-mass objects suggest that the rotation rate of young ($\sim$1-100~Myr) planetary-mass objects are well below their breakup rate \citep{Bryan_Benneke_Knutson_Batygin_Bowler_2018},
we consider that at some point the magnetic torque will play an important role in extracting the angular momentum from the protoplanet in the form of, for example, a magnetospheric wind as in the case of classical T Tauri stars (disk wind driven by the rotating protoplanetary magnetosphere). The shock heating around the CPD surfaces will help magnetic fields couple with the CPD gas by increasing the ionization degree.
However, we consider that there is a large uncertainty for the timing. Unlike the systems of classical T Tauri stars, the accretion occurs not only around the midplane but also in the polar regions in the case of proto-gas giants. Therefore, the magnetospheric wind can be confined by the polar accretion depending on the density of the polar accretion.
When the wind fails to blow, an efficient angular momentum loss may not occur because the wind gas will fall onto the CPD and come back to the protoplanet. As the density of the polar accretion will depend on the density of the parent protoplanetary disk, the spin evolution will be related to the evolution of the protoplanetary disk.
Detailed studies of the angular momentum exchange processes around protoplanets in evolving protoplanetary disks are required to reveal the spin evolution.

\acknowledgments
We would like to thank Haruhi Enomoto for fruitful discussions.
S.~T, A.~Y. and M.~I. received support from JSPS KAKENHI grant Nos. JP18K13579, JP17H01153 and JP18H05439.

\appendix

\section{Dependence of the spectral width on the accretion density}\label{sec:appendix-a}
\begin{figure}
\epsscale{0.5}
\plotone{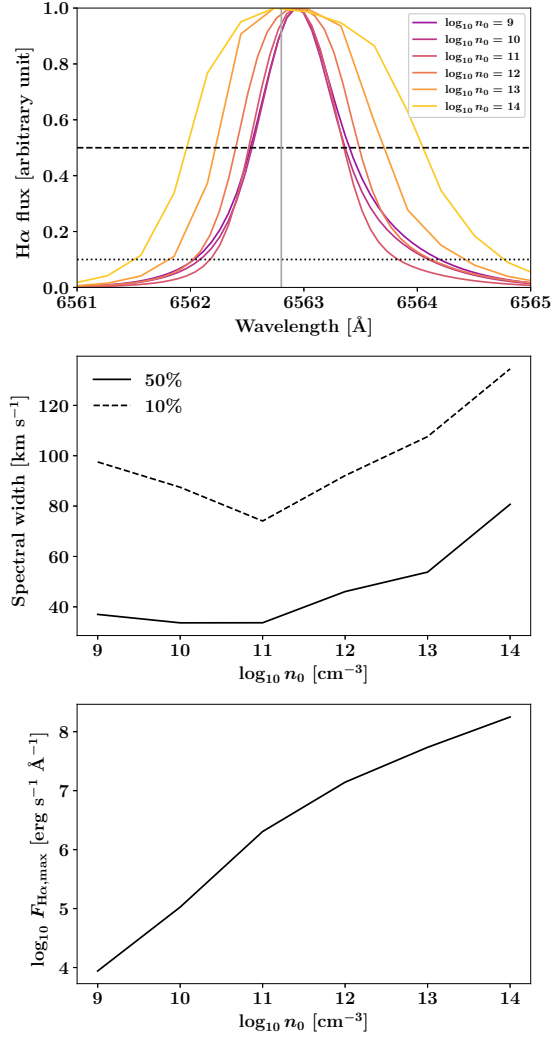}
\caption{Top: the H$\alpha$ line profiles for different preshock densities. The accretion speed is fixed to be $100~{\rm km~s^{-1}}$. The unit of the preshock number density ($n_0$) is ${\rm cm}^{-3}$. The horizontal dashed and dotted lines indicate the 50\% and 10\% full line widths, respectively. Middle: the dependence of the H$\alpha$ line broadening on the preshock density. The solid and dashed lines denote the 50\% and 10\% full line widths, respectively. Bottom: the dependence of the maximum H$\alpha$ line flux on the preshock density.
\label{fig:HaSED_den_dependence}}
\end{figure}

Using our 1D radiation-hydrodynamic code, we study how the line profile changes depending on the accretion density (preshock density), $n_0$. Here we fix the accretion speed to be 100~${\rm km~s^{-1}}$. The top panel of Figure~\ref{fig:HaSED_den_dependence} shows the line profiles for different pre-shock densities. The middle panel displays the spectral width as a function of the preshock density, where the 10\% full line width shows a nonmonotonic dependence. Below $n_0\approx 10^{11}~{\rm cm^{-3}}$, the spectral width decreases with the preshock density, while increasing with $n_0$ above that density. The H$\alpha$ line flux is a monotonic function, as it is nearly proportional to the kinetic energy flux of the accretion flow.

Below $n_0\approx 10^{11}~{\rm cm^{-3}}$, the spectral broadening at 10\% of the spectral peak decreases as the density increases. This is because the relative importance of the emission from the cooled hydrogen atoms to the emission from the strongly red-shifted, hot neutral hydrogen atoms becomes larger as the preshock density gets higher.
On the other hand, above $n_0\approx 10^{11}~{\rm cm^{-3}}$, the width at 10\% of the spectral peak increases because of the self-absorption of the H$\alpha$ radiation around the line center (see the right accretion column). The reduction of the H$\alpha$ radiation around the line center results in the apparent line broadening \citep{2018ApJ...866...84A, 2019ApJ...885L..29A}.

\section{Estimation of Cooling Time in the Post-CPD-shock region}~\label{sec:appendix-b}
\begin{figure}
\epsscale{0.9}
\plotone{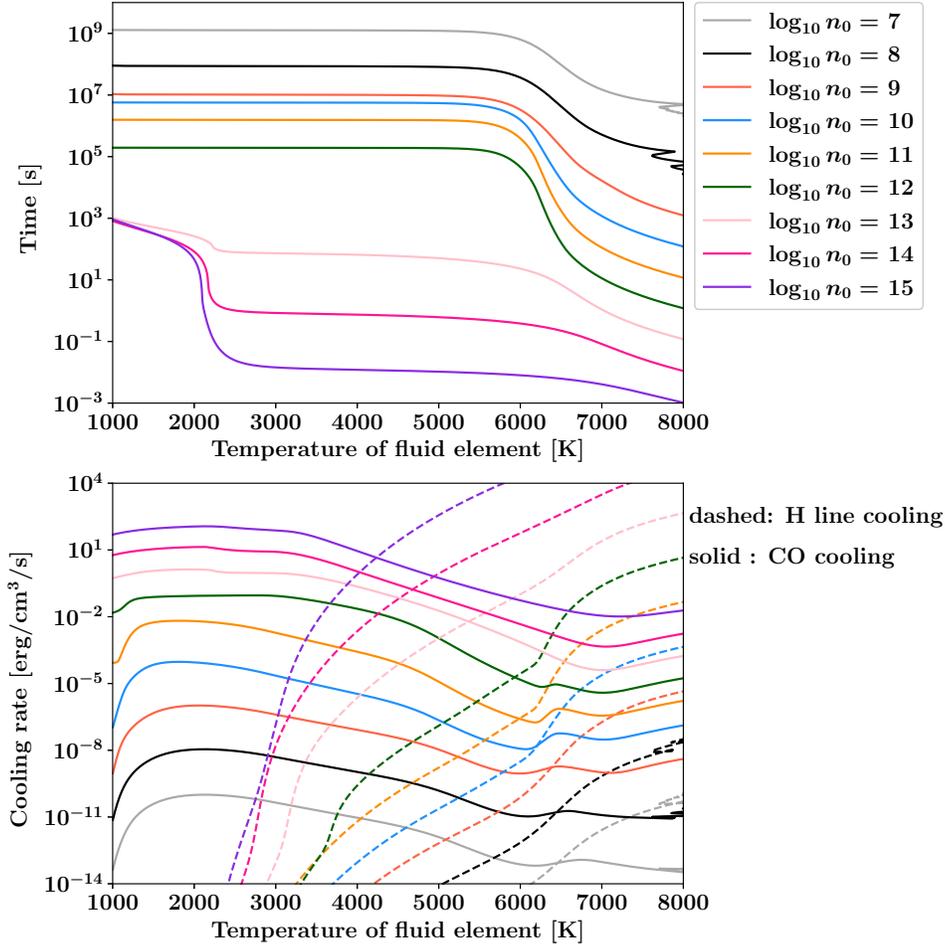}
\caption{Long-term ($\gg 1$~s) evolution of the post-CPD-shock gas with different preshock densities ($n_0$ in units of cm$^{-3}$). The accretion speed (or the preshock velocity) is fixed to be 40~km~s$^{-1}$. The lines are color-coded according to $n_0$ in both panels. Top: the elapsed time after the passage of the shock as a function of the temperature of a Lagrangian fluid element (we track the temperature of a Lagrangian fluid element). The vertical axis shows the elapsed time. Each value of the elapsed time at a temperature of 1000~K indicates the cooling timescale for each $n_0$. Bottom: the evolution of the cooling rate due to the H line cooling (dashed) and the CO cooling (solid) for a Lagrangian fluid element.
\label{fig:radiative_cooling}}
\end{figure}

We investigate how the gas cools down after the passage of the CPD shock. We consider the cooling of the gas in the region deeper than the H$\alpha$-emitting layer. The top panel of Figure~\ref{fig:radiative_cooling} shows the temporal change in temperature along the motion of a Lagrangian fluid element in the post-CPD-shock region for different preshock densities $n_0$. The vertical axis denotes the elapsed time after the passage of the shock. Each value of the elapsed time at a temperature of 1000~K indicates the cooling timescale for each $n_0$. Here we only show the temperature evolution after the temperature drops to 8000~K, although the temperature is several 10$^4$~K in the flow just behind the shock \citep[see, e.g., Fig.~2 of][]{2018ApJ...866...84A}. Almost no molecules exist just after the shock because of dissociation induced by shock heating, and the hydrogen line emission reduces the temperature of the shocked gas to $\sim 10^4$~K within $\sim 1$~s. Then, as shown in Figure~\ref{fig:radiative_cooling}, once the temperature becomes low enough that excited hydrogen is unavailable, the gas cooling timescale drastically increases due to the lack of effective coolant molecules such as CO, and the temperature stays at several $10^3$~K. After that, a sudden cooling toward $T=1000$~K occurs once a sufficient amount of CO molecules is formed (for example, see the behavior around $T\approx 6000$~K for $10^9~{\rm cm^{-3}}\lesssim n_0\lesssim 10^{12}~{\rm cm^{-3}}$). The local number density $n$ increases as the temperature decreases. The typical compression ratio is $\sim$100, which is comparable to the compression ratio across a strong shock for the gas with $\gamma=1.01$.

Thus, in the postshock region, the main cooling process switches from the H line cooling to the CO vibrational cooling as time proceeds. The bottom panel of Figure~\ref{fig:radiative_cooling} displays both cooling rates as functions of the temperature of a Lagrangian fluid element. Note that the figure shows the time-evolving cooling rate for a Lagrangian fluid element. For the cases of $n_0\lesssim 10^{12}~{\rm cm^{-3}}$, the transition occurs around $6000$~K.

In response to the CO cooling, the temperature of a Lagrangian fluid element evolves as
\begin{align}
\frac{de_{\rm int}}{dt} &\approx -\Lambda_{\rm CO(V)}\\
    \frac{dT}{dt}&\propto -\Lambda_{\rm CO(V)}n^{-1}\label{eq:dTdt},
\end{align}
where $e_{\rm int}$ is the internal energy density, $n$ is the local number density and $\Lambda_{\rm CO(V)}~[{\rm erg~cm^{-3}~s^{-1}}]$ is the cooling rate via the CO vibrational emission per volume. The functional form of $\Lambda_{\rm CO(V)}$ depends on the local density $n$ \citep[see][]{1993ApJ...418..263N,2001Icar..153..430I}. When $n$ is not sufficiently high to realize the thermal equilibrium for the rovibrational states of CO molecules, the vibrational emission efficiency is controlled by the collision of hydrogen atoms. Namely,
\begin{align}
    \Lambda_{\rm CO(V)} \propto n({\rm H})n({\rm CO}),
\end{align}
where $n({\rm X})$ is the number density of a chemical species X. We can rewrite this relation using the relative abundance, $y(X)=n(X)/n_{\rm H}$, where $n_{\rm H}$ is the hydrogen nuclei number density. Here we note that the local number density $n$ is comparable to $n_{\rm H}$ within a factor of 2. Then, we obtain
\begin{align}
    \Lambda_{\rm CO(V)}  \propto y({\rm H})y({\rm CO})n^2\label{eq:LambdaCO_low}
\end{align}
When $n$ is sufficiently high to realize the thermal equilibrium, we can assume the local thermodynamic equilibrium (LTE) and the cooling function behaves as
\begin{align}
    \Lambda_{\rm CO(V)}  \propto n({\rm CO}) \propto y({\rm CO})n.\label{eq:LambdaCO_high}
\end{align}
We find that the transition occurs around at $n_0\sim 10^{9}~{\rm cm^{-3}}$. By combining Equations~\ref{eq:dTdt}, \ref{eq:LambdaCO_low}
, and \ref{eq:LambdaCO_high}, we obtain the density dependence of the temperature decrease rate as
\begin{align}
    \bigg\rvert\frac{dT}{dt}\bigg\rvert \propto 
    \begin{cases}
        y({\rm CO})n & {\rm for}\phantom{a} n_0 \lesssim 10^{9}~{\rm cm^{-3}} \phantom{a} {\rm (non-LTE)}\\
        y({\rm CO}) & {\rm for}\phantom{a} n_0 > 10^{9}~{\rm cm^{-3}} \phantom{a} {\rm (LTE)}
    \end{cases}
\end{align}
Therefore, to estimate the cooling timescale, we need to understand the behavior of the abundance $y({\rm CO})$. As $y({\rm CO})$ is controlled by chemical reactions, we will see the relevant chemical reactions.
We note that when $n_0>10^{13}~{\rm cm^{-3}}$, the H line cooling dominates the CO cooling even around at $T=4000$--$5000$~K and the density dependence of the cooling timescale is distinct from that of the lower-density cases. The H line cooling accelerates the cooling, but it cannot be important below $T=4000$--$5000$~K. As a result, the cooling timescale is limited by the CO cooling and the cooling timescale will not be smaller than $\sim 10^3$~s even for the case with $n_0=10^{15}~{\rm cm^{-3}}$.

CO is mainly formed via the following process:
\begin{align}
    \mathrm{C+OH\longrightarrow CO+H.}\label{eq:reaction-co}
\end{align}
After the temperature becomes $\sim 6000$~K, we can ignore the destruction of CO molecules. Therefore, the rate of the change in the number density of CO molecules, $n({\rm CO})$, can be written as
\begin{align}
    \frac{dn({\rm CO})}{dt}&=K({\rm CO,form})n({\rm C})n({\rm OH}),
\end{align}
or
\begin{align}
    \frac{dy({\rm CO})}{dt}&=K({\rm CO,form})y({\rm C})y({\rm OH})n\label{eq:yCO},
\end{align}
where $K({\rm CO,form})$ is the reaction rate coefficient for Chemical Reaction~\ref{eq:reaction-co}. This suggests that the rate of the change in the CO abundance $y({\rm CO})$ is proportional to $y({\rm OH})$.

As the CO formation rate depends on $y({\rm OH})$, we consider the OH formation. The main formation and destruction reactions are found to be
\begin{align}
    \mathrm{O + H_2 \longrightarrow OH + H}\label{eq:OH-form}\\
    \mathrm{OH + H \longrightarrow O + H_2}\label{eq:OH-dest}
\end{align}
We found that the abundance $y({\rm OH})$ is determined via the balance between these reactions. Therefore, if we write the reaction rates of Chemical Reactions~\ref{eq:OH-form} and \ref{eq:OH-dest} as $K({\rm OH,form})$ and $K({\rm OH,dest})$, respectively, we get
\begin{align}
    \frac{dn({\rm OH})}{dt}&=K({\rm OH,form})n({\rm O}) n({\rm H_2})-K({\rm OH,dest})n({\rm OH}) n({\rm H})=0\\
    y({\rm OH})&=\frac{K({\rm OH,form})}{K({\rm OH,dest})}\frac{y({\rm O}) y({\rm H_2})}{y({\rm H})},\label{eq:yOH}
\end{align}
which suggests that $y({\rm OH})$ is proportional to $y({\rm H_2})$.

The main H$_2$ formation process is found to depend on the density. For the case of $n_0 \lesssim 10^{11}~{\rm cm^{-3}}\equiv n_{\rm c0}$, H$_2$ molecules are formed through the two-body reaction via H$^{-}$:
\begin{align}
    \mathrm{H+e^{-} \longrightarrow H^- + \gamma}\\
    \mathrm{H^{-}+H \longrightarrow H_2 + e^-}
\end{align}
The first reaction limits the reaction rate. We will write the reaction rate of the first reaction as $K^{(1)}({\rm H_2,form})$.
We assume that the following two-body reaction on dust grains is unimportant because of the dust destruction by the shock:
\begin{align}
  \mathrm{H+H + grain \longrightarrow H_2}  
\end{align}
When $n_0> n_{\rm c0}$, H$_2$ molecules are formed via the three-body reaction:
\begin{align}
    \mathrm{H+H+H\longrightarrow H_2+H}
\end{align}
We will write the reaction rate as $K^{(2)}({\rm H_2,form})$. H$_2$ molecules are destructed through the two-body reaction in both cases:
\begin{align}
\mathrm{H_2+H \longrightarrow H+H+H}
\end{align}
The reaction rate will be denoted as $K({\rm H_2,dest})$. In both cases, $y({\rm H_2})$ is determined by the balance between the formation and destruction reactions listed above. In short, $y({\rm H_2})$ can be expressed as
\begin{align}
    y({\rm H_2})=
    \begin{cases}
    \frac{K^{(1)}({\rm H_2,form})}{K({\rm H_2,dest})}y({\rm H}) &{\rm for}\phantom{a} n_0\lesssim n_{\rm c0}\\
    \frac{K^{(2)}({\rm H_2,form})}{K({\rm H_2,dest})}y({\rm H})^2 n &{\rm for}\phantom{a} n_0> n_{\rm c0}
    \end{cases}\label{eq:yH2}
\end{align}
We note that $y({\rm H})\approx 1$ and can be regarded as a constant above $T\approx 10^3$~K before H$_2$ starts to form.

By combining Equations~\ref{eq:yCO}, \ref{eq:yOH}, and \ref{eq:yH2}, we find how the increasing rate of the CO abundance depends on the local density $n$:
\begin{align}
    \frac{dy({\rm CO})}{dt}&\propto y({\rm OH})n\propto y({\rm H_2})n\\ 
    &\propto 
    \begin{cases}
    n & {\rm for}\phantom{a} n_0\lesssim n_{\rm c0}\\
    n^2 & {\rm for}\phantom{a} n_0> n_{\rm c0}
    \end{cases}
\end{align}
As $\mathrm{H}_2$ and OH form much faster than CO, the $\mathrm{H_2}$ and OH abundances are functions of the local temperature and density and do not explicitly depend on time.

Table~\ref{tab:cooling} summarizes the radiative cooling process in the CPD postshock region for the case of the preshock velocity of 40~${\rm km~s^{-1}}$. As the CO cooling rate depends on the CO abundance $y({\rm CO})$, we also list the density dependence of the CO formation rate. The dependence of the cooling rate on the local density $n$ changes depending on whether LTE is realized for the rovibrational states of CO molecules.
The density dependence of the CO formation rate also shows a transition at a different $n_0$, which reflects the fact that the main H$_2$ formation process switches from the two-body to three-body reactions.
In our 2D hydrodynamic model, the preshock density of CPD shocks is $n_0\sim 10^9~{\rm cm^{-3}}$, similar to the density at which a transition of the main cooling process (non-LTE to LTE CO cooling) occurs (also see Figure~\ref{fig:accretion-structure}). Figure~\ref{fig:radiative_cooling} suggests the cooling timescale of $10^7$~s. When $n_0=10^8~{\rm cm^{-3}}$, the cooling time becomes longer by an order of magnitude because $|dT/dt|$ is proportional to the density. When $n_0 > 10^9~{\rm cm^{-3}}$, the cooling timescale becomes smaller but it is not inversely proportional to the density. The cooling timescale implicitly depends on $n$ through $y({\rm CO})$ and $n$ changes with time, which results in a complicated behavior. 

\begin{deluxetable*}{ccccc}
\tablenum{1}
\tablecaption{Summary of the radiative cooling process in the CPD post-shock region (for the case of the pre-shock velocity of 40~${\rm km~s^{-1}}$)\label{tab:cooling}}
\tablewidth{0pt}
\tablehead{
\colhead{Pre-shock density} & \colhead{Dominant cooling process} & \colhead{Dependence of cooling rate} & \colhead{H$_2$ formation process} & \colhead{Dependence of CO formation rate}  \\
\colhead{$n_0~[\rm cm^{-3}]$} & \colhead{} & \colhead{$\displaystyle \bigg \rvert \frac{dT}{dt}\bigg \rvert~[\rm K~s^{-1}]$} & \colhead{} & \colhead{$\displaystyle \frac{dy({\rm CO})}{dt}\propto y({\rm OH})n\propto y({\rm H_2})n$} 
}
\startdata
$n_0 \lesssim 10^9$ & CO cooling, non-LTE & $\propto y({\rm CO})n$ & two-body reaction via H$^{-}$ & $\propto n$\\
$10^9 < n_0 \lesssim 10^{11}$ & CO cooling, LTE & $\propto y({\rm CO})$ & two-body reaction via H$^{-}$ & $\propto n$\\
$10^{11}< n_0 \lesssim 10^{13}$ & CO cooling, LTE & $\propto y({\rm CO})$ & three-body reaction & $\propto n^2$\\
$10^{13}<n_0$ & H line cooling & --- & --- & ---\\
\enddata
\tablecomments{$y(\rm X)$ denotes the relative abundance of a chemical species X. $n$ is the local number density in the CPD postshock region. Here we consider the cooling of the gas in the region deeper than the H$\alpha$-emitting layer.}
\end{deluxetable*}

\bibliography{ProtoGasGiantAccShock01}{}
\bibliographystyle{aasjournal}

\end{document}